\documentclass[fleqn,usenatbib]{mnras}

\usepackage{newtxtext,newtxmath}
\usepackage{multirow}
\usepackage{graphicx}   
\usepackage{amsmath}    

\usepackage{amssymb}    
\usepackage{threeparttable}
\usepackage{array}
\usepackage[figuresright]{rotating}
\usepackage{ulem}
\usepackage{hyperref}
\usepackage{cleveref}

\usepackage[T1]{fontenc}

\DeclareRobustCommand{\VAN}[3]{#2}
\let\VANthebibliography\thebibliography
\def\thebibliography{\DeclareRobustCommand{\VAN}[3]{##3}\VANthebibliography}

\usepackage{graphicx}	
\usepackage{amsmath}	
\usepackage{amssymb}	
\usepackage{multirow}

\newcommand{\oiii}{[O\,{\sc iii}]}
\newcommand{\oii}{[O\,{\sc ii}]}
\newcommand{\hb}{H$\beta$}
\newcommand{\ha}{H$\alpha$}
\newcommand{\nii}{[N\,{\sc ii}]}
\newcommand{\sii}{[S\,{\sc ii}]}
\newcommand{\greenpea}{{\it Green pea}}
\newcommand{\blue}{{\it Blueberry}}
\newcommand{\purple}{{\it Purple grape}}

\title[Double Peaked Green Peas]{Discovery of Five Green Pea Galaxies with Double-peaked Narrow [O\,{\sc iii}] Lines}

\author[Lin et al.]{
Ruqiu Lin,$^{1,2}$
Zhen-Ya Zheng,$^{1}$\thanks{Contact e-mail: \href{mailto:zhengzy@shao.ac.cn}{zhengzy@shao.ac.cn}}
Jun-Xian Wang,$^{3}$
Fang-Ting Yuan,$^{1}$
James E. Rhoads,$^{4}$
Sangeeta Malhotra,$^{4}$
\newauthor{
Tao An,$^{1,2,5}$
Chunyan Jiang,$^{1}$
Shuairu Zhu,$^{1,2}$
P.T. Rahna,$^{1}$
Xiang Ji,$^{1}$
Mainak Singha$^{6}$
}
\\
$^{1}$Shanghai Astronomical Observatory, Chinese Academy of Sciences, 80 Nandan Road, Shanghai 200030, China\\
$^{2}$School of Astronomy and Space Sciences, University of Chinese Academy of Sciences, No. 19A Yuquan Road, Beijing 100049, China\\
$^{3}$CAS Key laboratory for Research in Galaxies and Cosmology, Department of Astronomy, University of Science and Technology of China, Hefei, Anhui 230026, China\\
$^4$Astrophysics Science Division, NASA Goddard Space Flight Center, 8800 Greenbelt Road, Greenbelt, Maryland, 20771, USA\\
$^5$Key Laboratory of Radio Astronomy and Technology, Chinese Academy of Sciences, A20 Datun Road, Chaoyang District, Beijing, 100101, P. R. China \\
$^6$ Department of Physics \& Astronomy, University of Manitoba, 30A Sifton Road, Winnipeg, MB R3T 2N2, Canada\\
}

\date{Accepted 2023 June 13. Received 2023 June 5; in original form 2023 April 17}

\pubyear{2022}

\begin{document}
\label{firstpage}
\pagerange{\pageref{firstpage}--\pageref{lastpage}}
\maketitle

\begin{abstract}
Although double-peaked narrow emission-line galaxies have been studied extensively in the past years, only a few are reported with the \greenpea\ galaxies (GPs).
Here we present our discovery of five GPs with double-peaked narrow \oiii\ emission lines, referred to as DPGPs, selected from the LAMOST and SDSS spectroscopic surveys. 
We find that these five DPGPs have blueshifted narrow components more prominent than the redshifted components, with velocity offsets of \oiii$\lambda 5007 \AA$ lines ranging from 306 to 518 $\rm km\, s^{-1}$ and full widths at half maximums (FWHMs) of individual components ranging from 263 to 441 $\rm km\, s^{-1}$. By analyzing the spectra and the spectral energy distributions (SEDs), we find that they have larger metallicities and stellar masses compared with other GPs. The \ha\ line width, emission-line diagnostic, mid-infrared color, radio emission, and SED fitting provide evidence of the AGN activities in these DPGPs. They have the same spectral properties of Type 2 quasars.
Furthermore, we discuss the possible nature of the double-peaked narrow emission-line profiles of these DPGPs and find that they are more likely to be dual AGN. 
These DPGP galaxies are ideal laboratories for exploring the growth mode of AGN in the extremely luminous emission-line galaxies, the co-evolution between AGN and host galaxies, and the evolution of high-redshift galaxies in the early Universe. 

\end{abstract}




\section{Introduction}
\label{sec:intro}
\greenpea\ (GPs) galaxies are small and compact galaxies with strong \oiii\ emission lines at low-redshifts, and are firstly discovered by citizen scientists~\citep{Cardamone2009}. GPs are quite rare at low-redshifts, whereas they are widely believed to have been quite common in the early Universe \citep{Izotov2011}. Recently JWST has confirmed the existence of GPs at very high redshift \citep{Schaerer2022,Trump2022,Taylor2022,Rhoads2023,Curti2023}.
The huge distances of galaxies in the early Universe prevent us from studying these objects directly as they are very faint, but GPs at a lower redshift offer the chance to understand the earliest galaxies by studying their analogs in the nearby Universe.

Several authors had searched for GPs from the Sloan Digital Sky Survey \citep[SDSS,][]{Kollmeier2017} using color-selection and spectroscopic confirmation methods. \cite{Cardamone2009} firstly built a sample of 251 GP candidates by the SDSS {\it gri} colors, and confirmed $\sim$ 100 GPs with their SDSS spectra. However, the {\it gri} color-selection method is restricted to the redshift range from 0.112 to 0.360.
A more straightforward way for finding GPs is to directly select luminous \oiii\ emission lines from the spectroscopic surveys. 
For example, \cite{Izotov2011} searched for GPs using both imaging and spectroscopic data and obtain 803 luminous compact emission-line galaxies at 0.02 < {\it z} < 0.63. The GP surveys are also carried out with the Large Sky Area Multi-Object Fiber Spectroscopic Telescope \citep[LAMOST, ][]{Wang1996}. 
A catalog  from LAMOST DR9 of 1547 strong \oiii$\lambda 5007$ emission-line galaxies at 0 < {\it z} < 0.59 was published by \cite{Liu2022}.

To further bridge the connections between the earliest galaxies and the main populations of galaxies in the modern Universe, it is worth checking whether some of the GPs are galaxy mergers. Galaxy mergers can trigger star formation in the merger center \citep{Kennicutt1984,Patton2011} and the activity of the SMBHs of the merging galaxies \citep{Alonso2007,Ellison2011}.

To study galaxy mergers, a lot of work have been done with double-peaked narrow emission-line galaxies in the past decades\citep[e.g.][]{Ge2012, Wang2019,Maschmann2020}. However, these 
double-peaked narrow emission-line galaxies could also 
be triggered by binary supermassive black holes \citep[SMBHBs, e.g.,][]{Liu2010}, dual AGN \citep[DAGN, e.g.][]{Comerford2009, Wang2009,Ge2012,Muller-Sanchez2015,Liu2018}, and furthermore, the outflows or rotating disks~\citep[e.g.][]{Liu2010,Shen2011,Nevin2016,Wang2019,Maschmann2023}.

It is challenging to clearly distinguish DAGN from gas kinematics (outflow and rotating disk) when probing the nature behind the double-peak line profile.
Fiber spectroscopy alone cannot determine the nature of the double-peak features of these objects, because there is no spatial information inside the area covered by the spectroscopic fiber. Observations with high spatial resolution are required to confirm DAGN.
For example, several DAGN candidates with double-peaked narrow-line emission lines are only confirmed using HST imaging observation \citep{Comerford2009}, Very Long Baseline Interferometry \citep[VLBI, e.g.][]{Muller-Sanchez2015}, or long slit spectroscopy \citep[e.g.][]{Comerford2012,Muller-Sanchez2015}.


Here in this work, we report the discovery of 5 GPs with double-peaked narrow \oiii\ emission-lines from a sample of 1635 GPs selected from LAMOST and SDSS.
We perform the multi-wavelength analysis to investigate the physical properties of this sample. The data, the spectral analysis, and the sample selection are described in \S \ref{sec:sampleSelection}. We analyse the galaxy properties and AGN activities in \S \ref{sec:gal} and \S \ref{sec:agn}, respectively.
In \S \ref{sec:discussion}, we discuss the nature of our sample, and compare our sample with other samples of double-peaked emission-line galaxies. Throughout this paper, we assume the cosmological parameters of ${\rm H0 = 70\, 
km\,s^{-1}\,Mpc^{-1},\, \Omega_m = 0.3,\, and\ \Omega_\Lambda = 0.7}$.





\section{Data and Sample Selection}
\label{sec:sampleSelection}
\subsection{Parent Sample}
Our parent sample is composed of 1635 strong \oiii\ emission-line galaxies (ELGs) collected from two catalogs (\citealt{Liu2022} and \citealt{Cardamone2009}) selected from the LAMOST\footnote{http://www.lamost.org/} and SDSS\footnote{https://www.sdss4.org/} spectroscopic surveys, respectively.
The first catalog is mainly the input catalog of a PI project (PI: Junxian Wang) of the LAMOST extra-galactic survey add-on program. This sample is presented in \cite{Liu2022} for a total of 1547 strong \oiii\ emission line compact galaxies (1694 spectra from LAMOST DR9). 
This catalog contains \blue, \greenpea\,(GP), and \purple\ galaxies selected by SDSS {\it ugriz} colors and covering the redshift range from {\it z} = 0 to {\it z} = 0.72.
The second catalog is the GP sample of \cite{Cardamone2009}, including publicly available 80 star-forming galaxies and 8 narrow-line Seyfert 1 galaxies (NLS1) in the redshift range of 0.112 $<\,z\,<$ 0.360. These 88 objects are selected via SDSS {\it gri} colors and have SDSS spectroscopic observations. 

Most galaxies of our parent sample are unresolved in SDSS images with a small size ({\it petrorad\_r} $\leq\ 5\arcsec$ for LAMOST sample and {\it petrorad\_r} $\leq\ 2\arcsec$ for SDSS sample), as the radius sizes ({\it r} band) of these galaxies are close to the median point spread function (PSF) size of SDSS {\it r} band (see Figure 4 in \citealt{Liu2022}). Therefore, given the fiber diameters of 3\farcs3 and 3\arcsec for the LAMOST and SDSS spectra, respectively, the spectral properties of the entire galaxies of this sample are probed.


\begin{figure}
    \centering
    \includegraphics[width=0.5\textwidth]{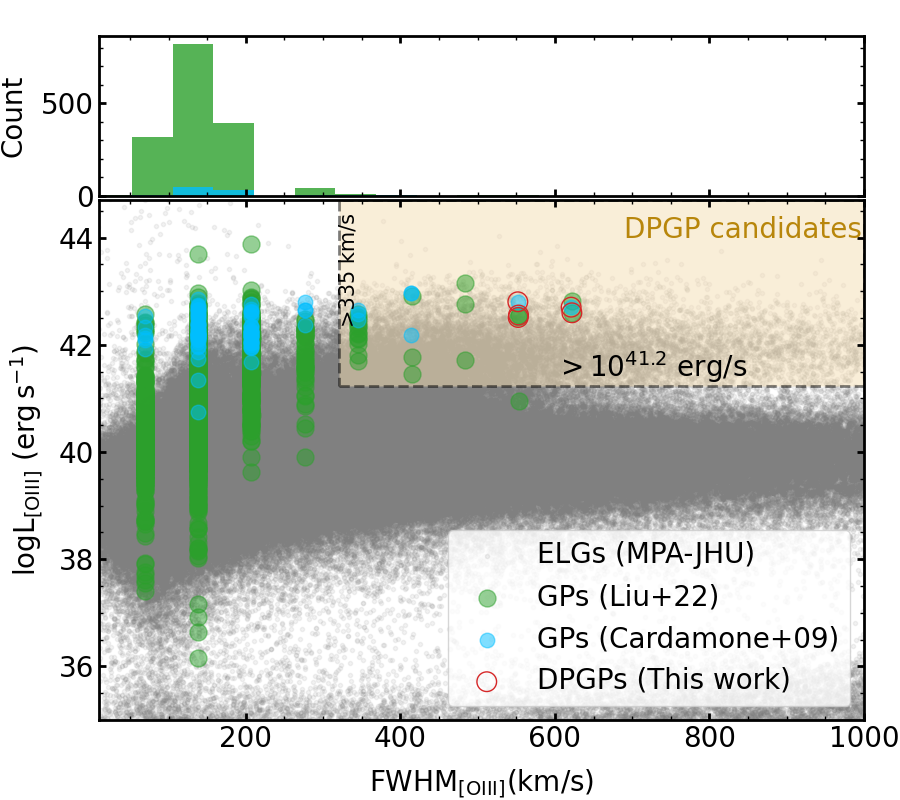}
    \caption{Luminosity and FWHM of the \oiii\ lines for GPs and emission-line galaxies. The green and blue circles represent the GP samples selected from LAMOST by \citet{Liu2022} and SDSS by \citet{Cardamone2009}, respectively. The DPGPs are marked in open red circles. The yellow shaded area represents the selection criteria for DPGP candidates. The emission-line galaxies in SDSS DR8 measured by MPA-JHU team are plotted as grey dots for comparison. All these galaxies have S/N(\oiii $\lambda\, 5007$)  $>$ 5 at its peak. The histogram of FWHM$\rm _{[OIII]}$ of our parent GP sample is also presented in the top panel.}
    \label{fig:LvsFWHM}
\end{figure}

\subsection{DPGP Candidates}
\label{DPGP_can}
The GP sample is characterized by the strong \oiii\ emission line, which is an advantage for searching the double-peaked line profile. To pre-select the GP candidates with double-peaked narrow line profile (DPGPs), we perform direct measurements to \oiii$\lambda\,5007\AA$ emission line for each galaxy. 
Before the measurements, we drop the spectra with no redshift measured from the LAMOST data reduction pipeline and 1769 spectra are left for the following analyses.
Because the original fluxes of LAMOST spectra are not physical, we follow \cite{Wang2018} and \cite{Liu2022} to re-calibrate the fluxes by comparing the magnitude of SDSS photometry and the re-calibrated fluxes are used throughout this paper.
To avoid the uncertainty introduced by line modeling with a single Gaussian profile, we directly measure the integrated line flux and the full width at half maximum (FWHM) from continuum-subtracted spectra. We note that this would introduce discrete FWHM values. We require the sample with the signal-to-noise ratio (S/N) larger than 5 at the peak of \oiii $\lambda\, 5007$ line to eliminate the contamination of the sky background. There are 1622 GPs (1732\footnote{There are 110 GPs with multiple spectra.} spectra) eventually measured.

We present the distribution of line luminosity versus line width of \oiii$\lambda5007\AA$ for our parent sample in Figure~\ref{fig:LvsFWHM}. 
We also plot the MPA-JHU SDSS DR8 galaxy sample\footnote{https://www.sdss4.org/dr17/spectro/galaxy\_mpajhu/} for comparison. Our parent sample is on the bright end compared to the MPA-JHU sample.
The LAMOST and SDSS spectra almost have the same spectral resolution of $\sim$ 1800 \citep{Du2016,York2000}, corresponding to an instrumental dispersion of $\rm \Delta V\sim69\, km\,s^{-1}$ per pixel at 5007 \AA. 
Most of this sample is concentrated in the diagram 
around the average FWHM\footnote{Throughout this paper, we did not correct instrument broadening in the FWHM or velocity dispersion we used, as the instrument dispersion only causes a small broadening (< 24 $\rm km\,s^{-1}$) in the integrated \oiii\ lines of our DPGPs.} value of $\rm FWHM_{ave} = 149\ km\,s^{-1}$ and with a standard deviation of $\rm \sigma_{FWHM} = 62\ km\,s^{-1}$. However, we can find a minority sample existed at the larger FWHM end. Since DPGPs often show broader line widths compared to normal GP galaxies, we set the following criteria to select DPGP candidates:
\begin{eqnarray}
    \rm FWHM_{[OIII]} & > & \rm FWHM_{ave}+3\sigma_{FWHM} = 335 \ km\,s^{-1}\\\nonumber
    \rm L_{[OIII]}\ & > & \rm 2.5\times10^{41}\,erg\,s^{-1},
\end{eqnarray}
where the minimum luminosity $L_{\rm[OIII]}$ is converted from the LAMOST limiting magnitude (19 mag) at {\it z} $\sim$ 0.72 (i.e., the maximum redshift in our GP sample) to further eliminate contamination from low S/N spectrum.
Only $\sim$ 2\% (29/1622) of GPs are selected as DPGP candidates. 


\subsection{Spectral Analysis and DPGP Sample}
We identify the DPGPs from DPGP candidates by modeling the double-peak profiles of their narrow \oiii\ emission lines. 
\label{sec:spectralAnalysis}
We model the emission lines by using the public Python spectroscopic analysis package \citep[{\it pyspeckit},][]{Ginsburg2011}. The package employs the Levenberg-Marquardt optimization method via the {\it mpfit} \citep{Markwardt2009} and {\it lmfit} \citep{Newville2014}.
We first shift the spectra to the rest frame with the redshift derived from LAMOST or SDSS data reduction pipeline. Since the starlight from the stellar population is more complicated for two kinematic systems like DPGPs, it is difficult to precisely model the emission and absorption of the stellar population. Thus, we simply compute the continuum in the nearby line-free regions. 

After subtracting the continuum, we fit \hb\ and \oiii\ emission lines, and also \ha\ and \nii\ lines if available, with Gaussian profiles. We use a single narrow component ($\sigma < 600\,\rm km\, s^{-1}$) for each line. 
Following \citealt{Hao2005}, we use two narrow components rather than one component if the reduced $\chi^2$ of the best fit is improved by at least 20\%. This criterion is based on the {\it F}-test and described as follows (see also the equation (5) in \citealt{Hao2005}): 
\begin{equation}
    f = \frac{(\chi_{1}^2-\chi_2^2)/(\textit{dof}_1-\textit{dof}_2)}{\chi_2^2/\textit{dof}_2},
\end{equation}
where the $\chi_1^2$\, ($\chi_2^2$) and $ \textit{dof}_1\, (\textit{dof}_2)$ are the reduced $\chi^2$ and the degrees of freedom of models of single-Gaussian and two-Gaussian profiles, respectively. 

All lines (\hb\ and \oiii\ lines, and also \ha\ and \nii\ lines if available) are fitted together. We fix the velocity offsets between the redshifted and blueshifted line systems and also fix the velocity shift of each line system (e.g., \hb\ and \ha\ as one system). For each line system, we require the same line widths in the fitting. We also add a broad component ($\sigma > 600\,\rm km\, s^{-1}$) into the model when the reduced $\chi^2$ with the new component can be improved by at least 20\%. We note that the velocity dispersion of the broad-line components of Balmer lines (i.e., \ha\ and \hb) is also fixed, because of the same excited mechanism. However, the broad components of the forbidden lines (i.e., \oiii\ and \oii) are independent.

We directly fit the narrow line width with the \oiii \ line. Although the line widths of narrow emission lines are routinely determined through the isolated \sii\ line, we note that the \sii\ line at $z \sim $ 0.3 is located at the edge of spectra with low S/N, which prevents us to determine the narrow line width from the \sii\ line. We also tie the line widths of broad \ha\ and \hb\ components (if existing) together.
Lastly, we adopt a laboratory ratio of about 3 for both \oiii$\lambda5007$/\oiii$\lambda 4959$ and \nii$\lambda 6583$/\nii$\lambda 6549$ \citep{Osterbrock2006}. 

\begin{figure*}
    \centering
    \includegraphics[width=0.38\textwidth]{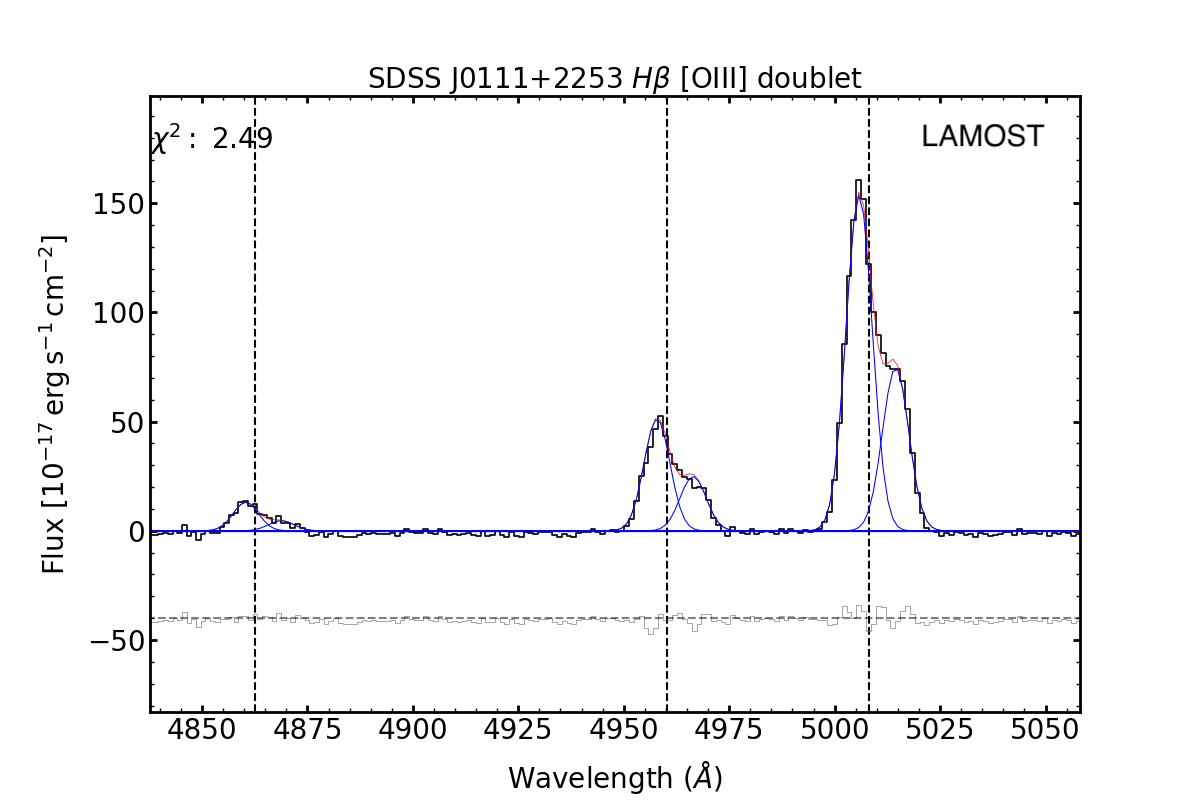}
    \includegraphics[width=0.38\textwidth]{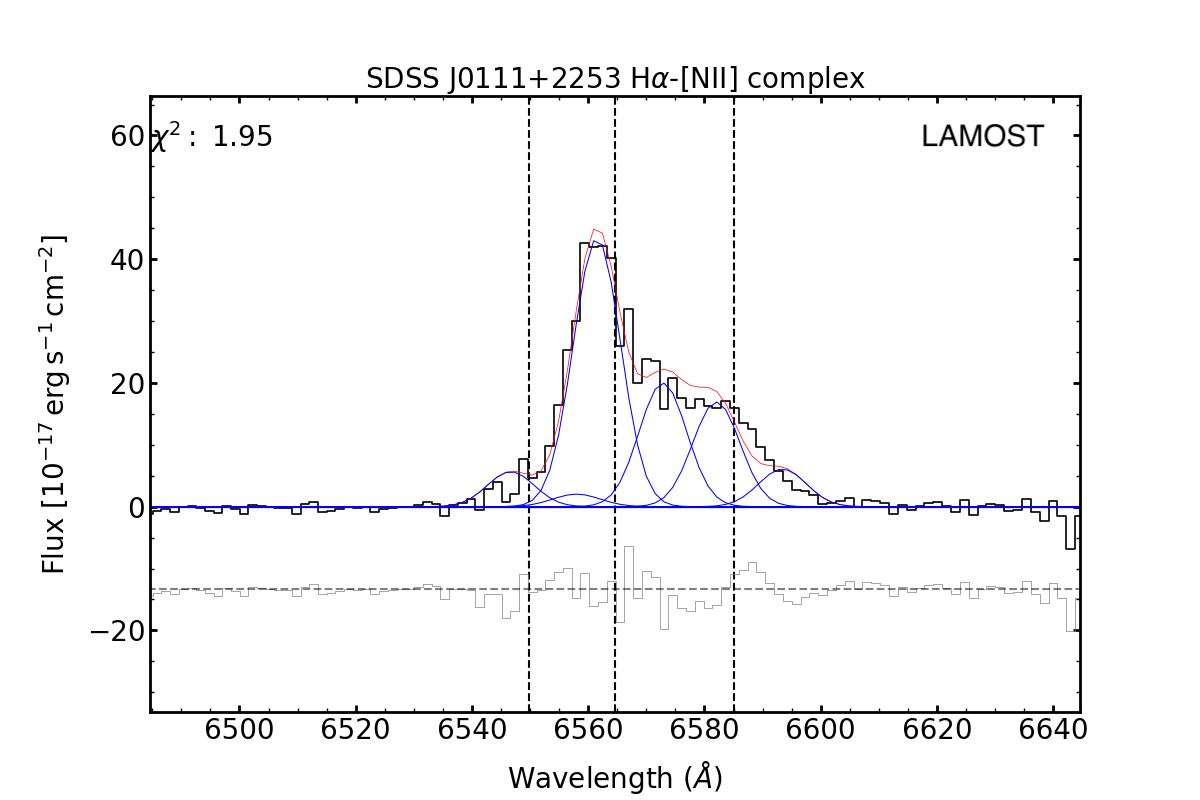}
    \includegraphics[width=0.38\textwidth]{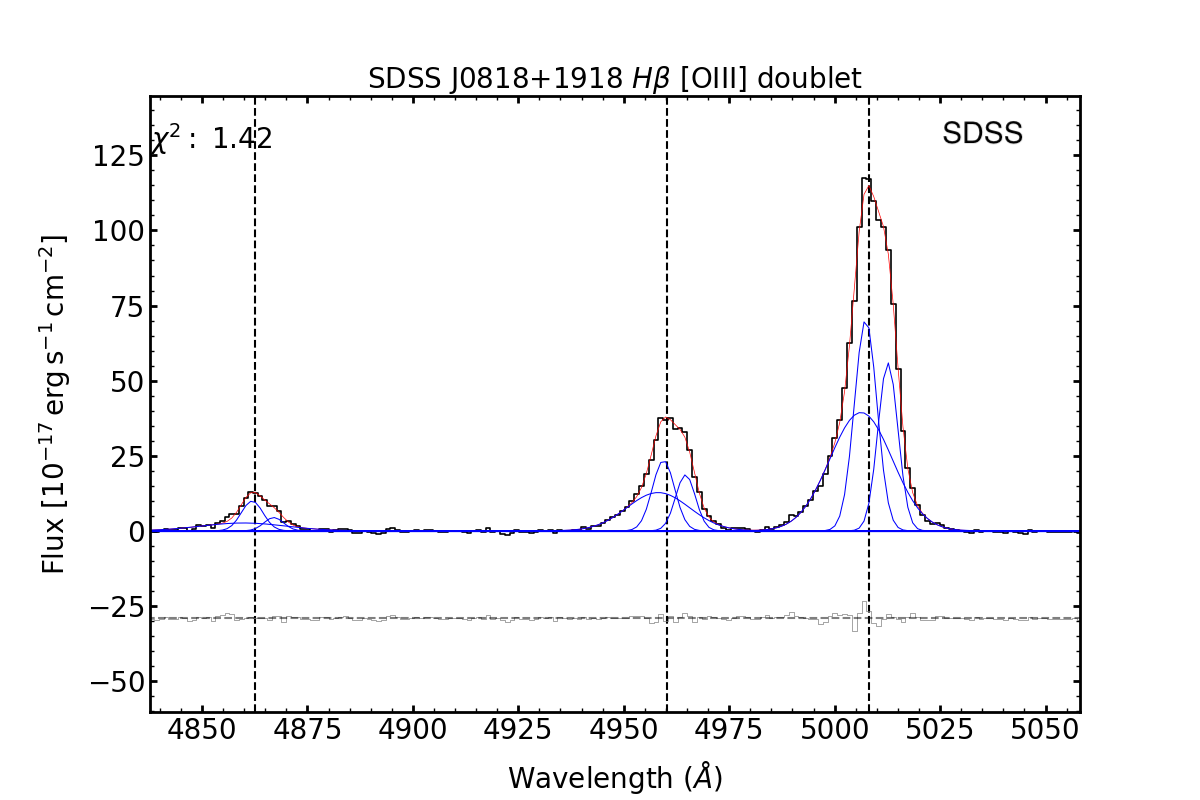}
    \includegraphics[width=0.38\textwidth]{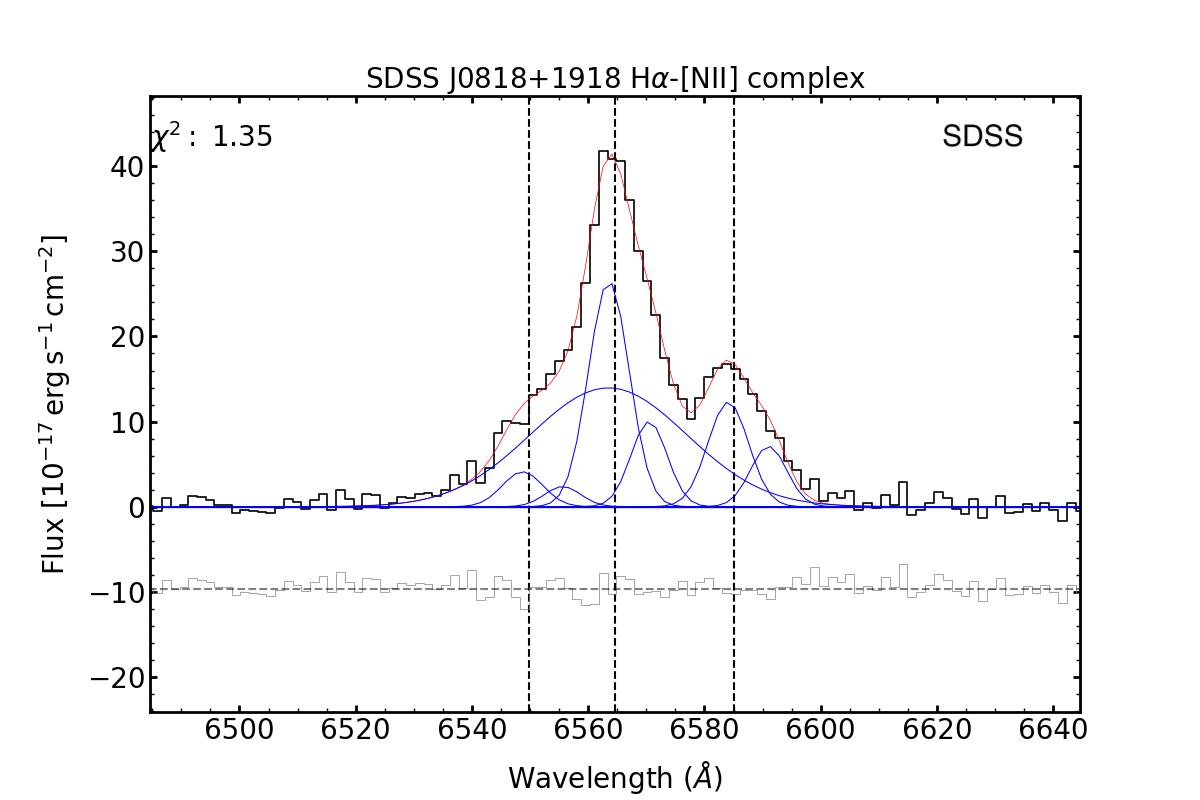}
    \includegraphics[width=0.38\textwidth]{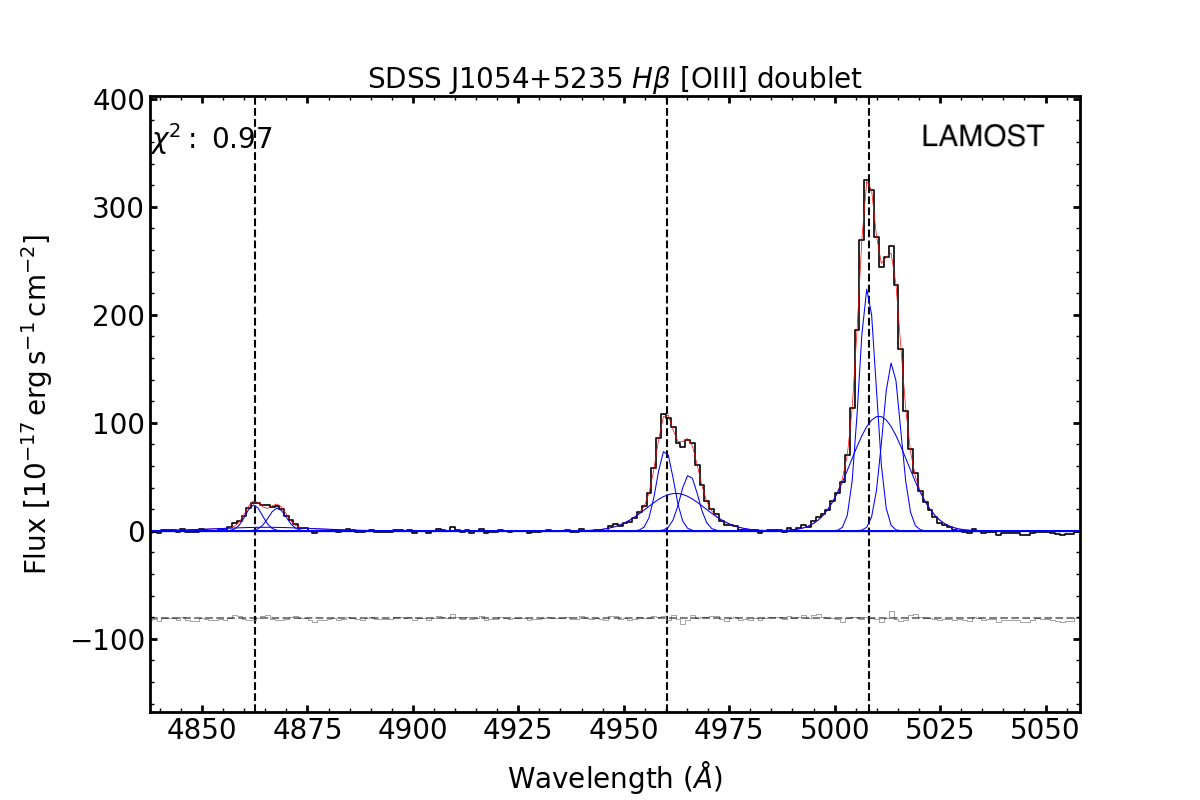}
    \includegraphics[width=0.38\textwidth]{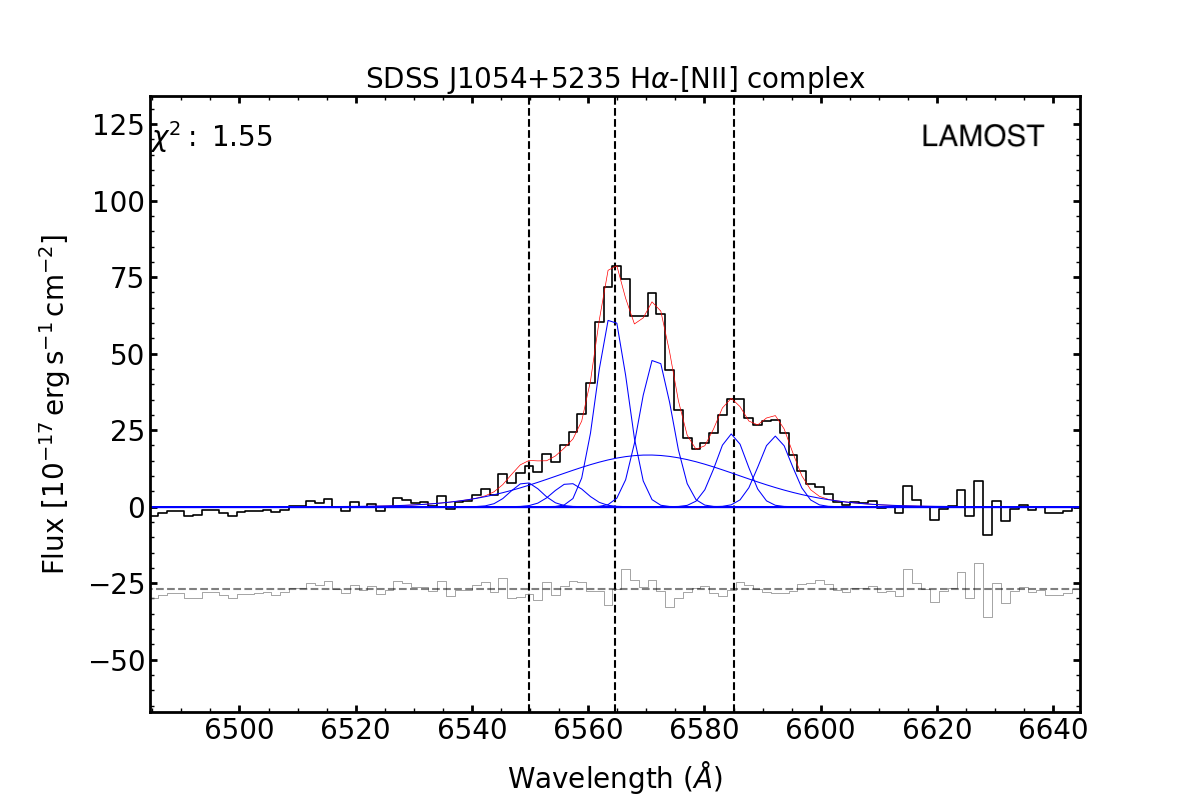}
    \includegraphics[width=0.38\textwidth]{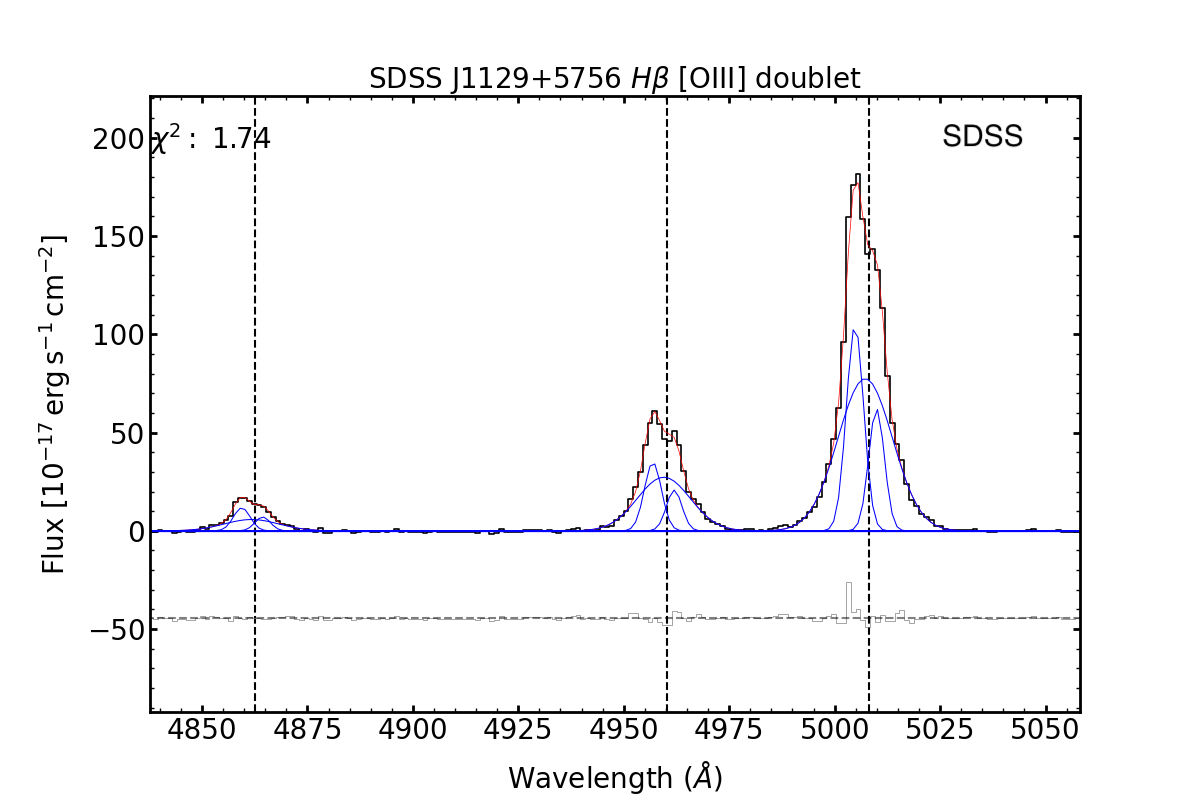}
    \includegraphics[width=0.38\textwidth]{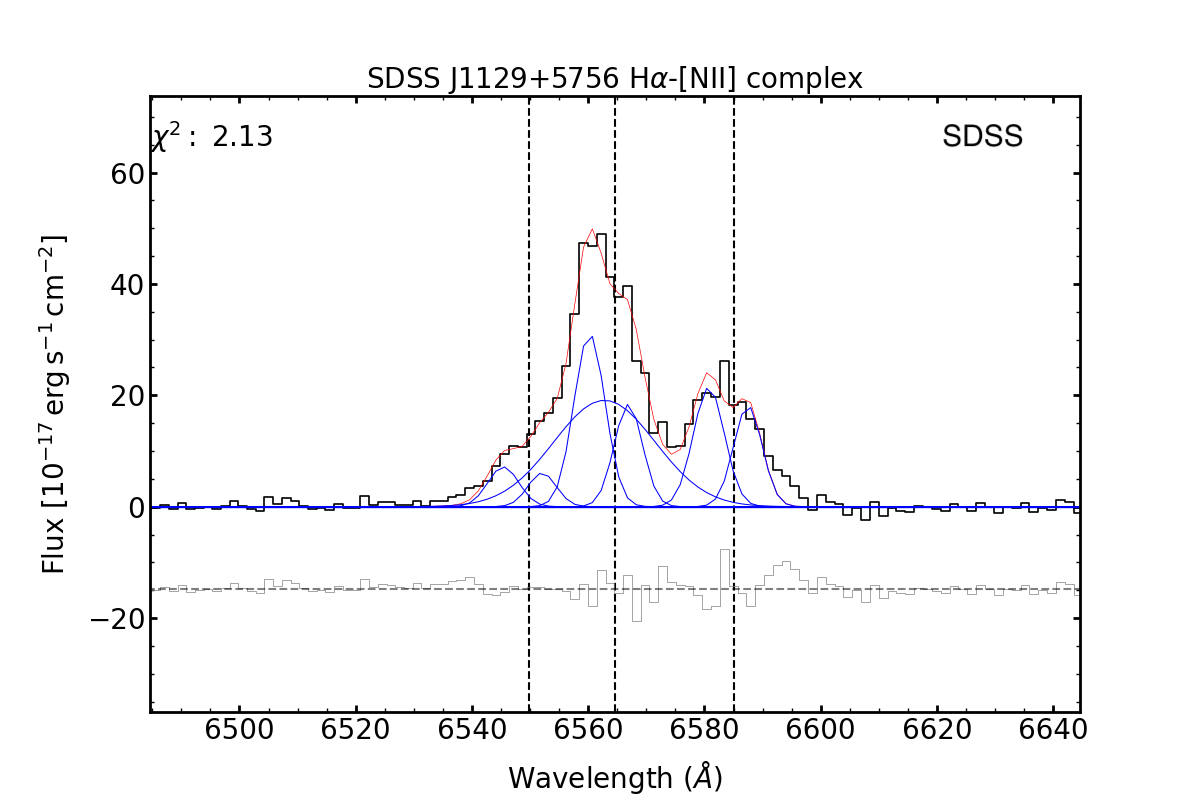}
    \includegraphics[width=0.38\textwidth]{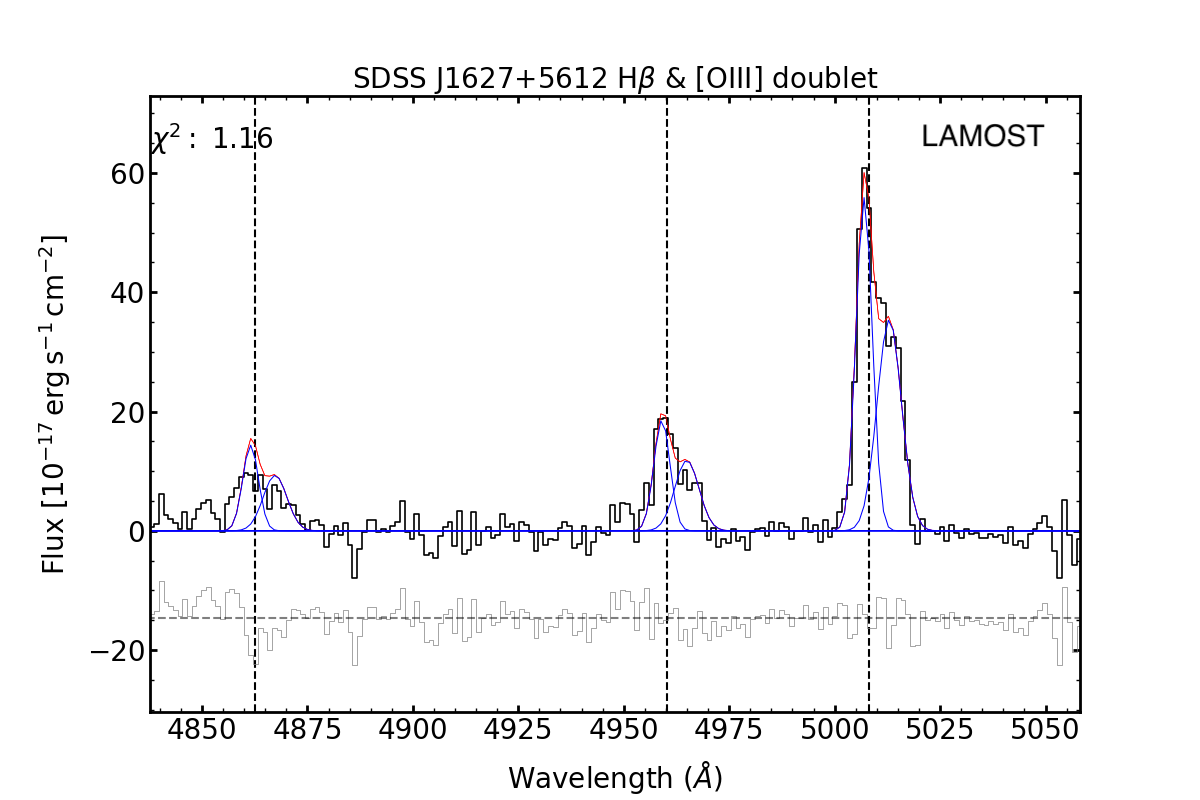}
    
    \caption{Emission line modeling for each of the 5 DPGPs. The black, red, and blue lines represent the observed spectra, composed model spectra, and  models for each component, respectively. The $\chi^2$ shown on the left top of each stamp is the reduced $\chi^2$ in the corresponding fitted wavelength region.The attributed data set of each spectrum is labeled in the upper right.}
    \label{fig:line_fitting}
\end{figure*}

As a result, there are 5 out of 29 DPGP candidates fitted well with two narrow components (of which 3 have broad components), whose basic information are listed in Table~\ref{tab:basicProperty}.  The detailed fitting results are shown in Table~\ref{tab:fittingResult} and Figure~\ref{fig:line_fitting}. 
Therefore, there are only $\sim$ 0.3\% (5/1622) of GPs selected as DPGPs.

\begin{table*}
\centering
\caption{The basic parameters of the 5 DPGPs.}
\label{tab:basicProperty}
\setlength{\tabcolsep}{3mm}{
\begin{tabular}{ccccccccc} 
\hline \hline
ID & SDSS Name & R.A. & Decl. & {\it z} & Velocity Offset (km/s) &
FWHM$_{\rm blue}$ (km/s)& FWHM$_{\rm red}$ (km/s)& 
Ref. \\ \hline
 
1 & J0111+2253 & 17.8888  & 22.8997 & 0.32 & 517.99 (2.73) & 437.64 (4.33) & 441.44 (8.47) &  L22\\
2 & J0818+1918 & 124.5008 & 19.3028 & 0.32 & 316.06 (5.46) & 376.54 (8.63) & 335.12 (10.88) &  C09 \\
3 & J1054+5235 & 163.7078 & 52.5882 & 0.32 & 343.61 (2.27) & 288.09 (4.92) & 305.53 (6.85)  & L22\\
4 & J1129+5756 & 172.2795 & 57.9348 & 0.31 & 305.59 (5.33) & 288.05 (9.20) & 262.90 (12.89) &C09 \\
5 & J1627+5612 & 246.9005 & 56.2072 & 0.49 & 357.35 (18.50) & 271.91 (12.71) & 408.33 (35.19) &L22\\
\hline
\end{tabular}
}
    \begin{tablenotes}
    \item NOTE. Column 1-5: Object ID, object name, Right Ascension, Declination, and redshift, respectively. Column 6: The velocity offsets between the blueshifted and redshifted components of \oiii$\lambda$5007 lines. Column 7 - 8: The FWHMs of the blueshifted and redshifted components of \oiii$\lambda$5007 lines, respectively. Column 9: L22 corresponds to \cite{Liu2022}, which utilized data from the LAMOST spectroscopic survey, while C09 corresponds to \cite{Cardamone2009}, which utilized data from the SDSS spectroscopic survey.
    The values in parentheses are 1-$\sigma$ errors.
    \end{tablenotes}
\end{table*}

\begin{table*}
\centering
\caption{The fitting parameters of emission lines for each of the 5 DPGPs.}
\label{tab:fittingResult}
\resizebox{\textwidth}{11.35cm}{
\setlength{\tabcolsep}{0.5cm}{
\begin{tabular}{lccccc}
\hline\hline
Lines & $\lambda_c$  & $V_c$  & FWHM  & Flux     & {Luminosity}   \\
  & $\rm \AA$ (rest frame)& $\rm km\,s^{-1}$ & $\rm km\,s^{-1}$ & $\rm 10^{-17}\,erg\,s^{-1}\,cm^{-2}$ & $\rm 10^{40}\,erg\,s^{-1}$ \\
\hline

\multicolumn{6}{c}{SDSS J0111+2253} \\

H$\beta_b$& 4860.39 (0.04) & -141.36 (2.27) & 437.64 (4.33) & 98.56 (4.51)    & 33.08 (1.50)  \\
H$\beta_r$& 4868.78 (0.00$^*$) & 376.63 (0.00$^*$)  & 441.44 (8.47) & 35.34 (2.97)    & 11.86 (0.99)  \\
\oiii$\lambda 4959_b$& 4957.88 (0.04) & -141.36 (2.23) & 437.64 (4.24) & 394.63 (8.10)   & 132.44 (2.70) \\
\oiii$\lambda 4959_r$& 4966.44 (0.00$^*$) & 376.63 (0.00$^*$)  & 441.44 (8.31) & 193.42 (7.12)   & 64.91 (2.37)  \\
\oiii$\lambda 5007_b$& 5005.81 (0.04) & -141.36 (2.21) & 437.64 (4.20) & 1195.34 (13.56) & 401.16 (4.52) \\
\oiii$\lambda 5007_r$& 5014.45 (0.07) & 376.63 (4.46)  & 441.44 (8.23) & 585.86 (12.55)  & 196.62 (4.18) \\
\nii$\lambda 6548_b$& 6546.68 (0.04) & -141.36 (1.69) & 437.64 (3.21) & 57.41 (7.17)    & 19.27 (2.39)  \\
\nii$\lambda 6548_r$& 6557.99 (0.00$^*$) & 376.63 (0.00$^*$)  & 441.44 (6.29) & 20.78 (4.72)    & 6.97 (1.57)   \\
\ha$_b$& 6561.52 (0.04) & -141.36 (1.68) & 437.64 (3.21) & 442.60 (11.34)  & 148.54 (3.78) \\
\ha$_r$& 6572.85 (0.00$^*$) & 376.63 (0.00$^*$)  & 441.44 (6.28) & 205.65 (2.92)   & 69.02 (0.97)  \\
\nii$\lambda 6583_b$& 6582.06 (0.04) & -141.36 (1.68) & 437.64 (3.20) & 173.15 (7.31)   & 58.11 (2.43)  \\
\nii$\lambda 6583_r$& 6593.43 (0.00$^*$) & 376.63 (0.00$^*$)  & 441.44 (6.26) & 62.67 (4.82)    & 21.03 (1.61)  \\

\multicolumn{6}{c}{SDSS J0818+1918} \\

H$\beta_b$& 4861.89 (0.09) & -48.33 (5.83)   & 376.54 (8.63)   & 64.33 (3.23)   & 22.41 (1.12)   \\
H$\beta_r$& 4867.02 (0.00$^*$) & 267.73 (0.00$^*$)   & 335.12 (10.88)  & 25.61 (2.46)   & 8.92 (0.85)    \\
H$\beta_{broad}$& 4859.81 (0.81) & -176.73 (50.11) & 1448.94 (73.79) & 65.30 (6.62)   & 22.75 (2.29)   \\
\oiii$\lambda 4959_b$& 4959.41 (0.09) & -48.33 (5.72)   & 376.54 (8.46)   & 155.33 (15.15) & 54.11 (5.24)   \\
\oiii$\lambda 4959_r$& 4964.64 (0.00$^*$) & 267.73 (0.00$^*$)   & 335.12 (10.67)  & 110.08 (11.92) & 38.35 (4.13)   \\
\oiii$\lambda 4959_{broad}$& 4958.16 (0.21) & -124.24 (12.46) & 1056.29 (25.25) & 236.83 (11.04) & 82.50 (3.82)   \\
\oiii$\lambda 5007_b$& 5007.36 (0.09) & -48.33 (5.66)   & 376.54 (8.38)   & 470.51 (18.20) & 163.91 (6.30)  \\
\oiii$\lambda 5007_r$& 5012.64 (0.12) & 267.73 (7.39)   & 335.12 (10.56)  & 333.43 (15.59) & 116.15 (5.39)  \\
\oiii$\lambda 5007_{broad}$& 5006.26 (0.11) & -114.06 (6.47)  & 1043.71 (12.31) & 730.65 (24.71) & 254.53 (8.55)  \\
\nii$\lambda 6548_b$& 6548.71 (0.09) & -48.33 (4.33)   & 376.54 (6.41)   & 35.89 (7.33)   & 12.50 (2.54)   \\
\nii$\lambda 6548_r$& 6555.61 (0.00$^*$) & 267.73 (0.00$^*$)   & 335.12 (8.08)   & 18.43 (3.48)   & 6.42 (1.21)    \\
\ha$_b$& 6563.56 (0.09) & -48.33 (4.32)   & 376.54 (6.39)   & 232.40 (11.24) & 80.96 (3.89)   \\
\ha$_r$& 6570.47 (0.00$^*$) & 267.73 (0.00$^*$)   & 335.12 (8.06)   & 78.13 (1.88)   & 27.22 (0.65)   \\
\ha$_{broad}$& 6563.50 (0.48) & -51.03 (21.77)  & 1448.94 (54.63) & 471.02 (28.90) & 164.09 (10.00) \\
\nii$\lambda 6583_b$& 6584.10 (0.09) & -48.33 (4.31)   & 376.54 (6.37)   & 108.26 (7.57)  & 37.71 (2.62) \\
\nii$\lambda 6583_r$& 6591.04 (0.00$^*$) & 267.73 (0.00$^*$)   & 335.12 (8.03)   & 55.60 (3.72)   & 19.37 (1.29)   \\

\multicolumn{6}{c}{SDSS J1054+5235} \\

H$\beta_b$& 4862.30 (0.04) & -23.25 (2.20)   & 287.76 (4.88)    & 119.36 (6.05)   & 41.62 (2.09)   \\
H$\beta_r$& 4867.87 (0.00$^*$) & 320.37 (0.00$^*$)   & 300.46 (6.92)    & 109.69 (6.40)   & 38.25 (2.21)   \\
H$\beta_{broad}$& 4865.11 (1.87) & 149.78 (115.52) & 1665.74 (105.43) & 94.25 (12.20)   & 32.87 (4.22)   \\
\oiii$\lambda 4959_b$& 4959.83 (0.04) & -23.25 (2.15)   & 287.76 (4.79)    & 379.14 (20.97)  & 132.21 (7.26)  \\
\oiii$\lambda 4959_r$& 4965.51 (0.00$^*$) & 320.37 (0.00$^*$)   & 300.46 (6.79)    & 274.87 (21.56)  & 95.85 (7.46)   \\
\oiii$\lambda 4959_{broad}$& 4962.34 (0.16) & 128.67 (9.92)   & 981.99 (26.31)   & 599.60 (32.67)  & 209.08 (11.30) \\
\oiii$\lambda 5007_b$& 5007.78 (0.04) & -23.25 (2.13)   & 287.76 (4.74)    & 1148.42 (27.67) & 400.46 (9.58)  \\
\oiii$\lambda 5007_r$& 5013.51 (0.05) & 320.37 (3.24)   & 300.46 (6.72)    & 832.58 (27.96)  & 290.32 (9.67)  \\
\oiii$\lambda 5007_{broad}$& 5010.50 (0.07) & 140.04 (4.20)   & 934.16 (14.77)   & 1763.37 (80.37) & 614.89 (27.81) \\
\nii$\lambda 6548_b$& 6549.26 (0.04) & -23.25 (1.63)   & 287.76 (3.63)    & 52.98 (8.31)    & 18.48 (2.88)   \\
\nii$\lambda 6548_r$& 6556.76 (0.00$^*$) & 320.37 (0.00$^*$)   & 300.46 (5.14)    & 53.74 (7.34)    & 18.74 (2.54)   \\
\ha$_b$& 6564.10 (0.04) & -23.25 (1.63)   & 287.76 (3.62)    & 421.57 (11.89)  & 147.00 (4.12)  \\
\ha$_r$& 6571.62 (0.00$^*$) & 320.37 (0.00$^*$)   & 300.46 (5.13)    & 343.61 (12.82)  & 119.82 (4.44)  \\
\ha$_{broad}$& 6570.35 (0.68) & 262.30 (31.03)  & 1665.74 (78.07)  & 655.69 (53.48)  & 228.64 (18.51) \\
\nii$\lambda 6583_b$& 6584.65 (0.04) & -23.25 (1.62)   & 287.76 (3.61)    & 159.81 (8.57)   & 55.73 (2.97)   \\
\nii$\lambda 6583_r$& 6592.19 (0.00$^*$) & 320.37 (0.00$^*$)   & 300.46 (5.11)    & 162.10 (7.82)   & 56.52 (2.71)   \\

\multicolumn{6}{c}{SDSS J1129+5756} \\

H$\beta_b$& 4859.44 (0.08) & -200.02 (4.87) & 288.05 (9.20)  & 58.09 (4.39)    & 18.57 (1.39)   \\
H$\beta_r$& 4864.39 (0.00$^*$) & 105.56 (0.00$^*$)  & 262.90 (12.89) & 32.03 (3.58)    & 10.24 (1.13)   \\
H$\beta_{broad}$& 4861.75 (0.35) & -57.28 (21.62) & 941.71 (71.72) & 93.46 (13.43)   & 29.87 (4.25)   \\
\oiii$\lambda 4959_b$& 4956.91 (0.08) & -200.02 (4.78) & 288.05 (9.02)  & 176.77 (15.94)  & 56.50 (5.05)   \\
\oiii$\lambda 4959_r$& 4961.96 (0.00$^*$) & 105.56 (0.00$^*$)  & 262.90 (12.63) & 95.97 (13.65)   & 30.68 (4.32)   \\
\oiii$\lambda 4959_{broad}$& 4959.48 (0.10) & -44.21 (5.78)  & 892.64 (14.93) & 428.43 (17.35)  & 136.93 (5.49)  \\
\oiii$\lambda 5007_b$& 5004.83 (0.08) & -200.02 (4.73) & 288.05 (8.93)  & 535.44 (22.44)  & 171.14 (7.11)  \\
\oiii$\lambda 5007_r$& 5009.93 (0.14) & 105.56 (8.27)  & 262.90 (12.51) & 290.71 (18.96)  & 92.91 (6.01)   \\
\oiii$\lambda 5007_{broad}$& 5007.25 (0.05) & -54.85 (3.14)  & 914.61 (10.14) & 1258.46 (42.33) & 402.23 (13.41) \\
\nii$\lambda 6548_b$& 6545.40 (0.08) & -200.02 (3.62) & 288.05 (6.83)  & 47.87 (8.85)    & 15.30 (2.80)   \\
\nii$\lambda 6548_r$& 6552.07 (0.00$^*$) & 105.56 (0.00$^*$)  & 262.90 (9.57)  & 36.97 (5.06)    & 11.82 (1.60)   \\
\ha$_b$& 6560.24 (0.08) & -200.02 (3.61) & 288.05 (6.82)  & 208.41 (14.03)  & 66.61 (4.44)   \\
\ha$_r$& 6566.92 (0.00$^*$) & 105.56 (0.00$^*$)  & 262.90 (9.55)  & 112.67 (10.74)  & 36.01 (3.40)   \\
\ha$_{broad}$& 6562.80 (0.41) & -83.03 (18.80) & 941.71 (53.13) & 418.95 (45.48)  & 133.90 (14.40) \\
\nii$\lambda 6583_b$& 6580.77 (0.08) & -200.02 (3.60) & 288.05 (6.79)  & 144.39 (9.45)   & 46.15 (2.99)  \\
\nii$\lambda 6583_r$& 6587.48 (0.00$^*$) & 105.56 (0.00$^*$)  & 262.90 (9.52)  & 111.50 (6.35)   & 35.64 (2.01)   \\

\multicolumn{6}{c}{SDSS J1627+5612} \\

H$\beta_b$& 4861.54 (0.12) & -70.44 (7.33)  & 271.91 (12.71) & 67.35 (8.15)   & 62.29 (7.44)   \\
H$\beta_r$& 4867.33 (0.00$^*$) & 286.91 (0.00$^*$)  & 408.33 (35.19) & 65.59 (9.18)   & 60.66 (8.38)   \\
\oiii$\lambda 4959_b$& 4959.05 (0.12) & -70.44 (7.18)  & 271.91 (12.46) & 89.10 (12.61)  & 82.41 (11.50)  \\
\oiii$\lambda 4959_r$& 4964.96 (0.00$^*$) & 286.91 (0.00$^*$)  & 408.33 (34.49) & 84.94 (11.53)  & 78.56 (10.52)  \\
\oiii$\lambda 5007_b$& 5006.99 (0.12) & -70.44 (7.11)  & 271.91 (12.34) & 269.89 (17.18) & 249.62 (15.67) \\
\oiii$\lambda 5007_r$& 5012.96 (0.29) & 286.91 (17.08) & 408.33 (34.16) & 257.29 (23.38) & 237.96 (21.33) \\

\hline
\end{tabular}}}
    \begin{tablenotes}
    \item NOTE. From left to right, there are the emission line species, the line center, the velocity offset from zore velocity derived with the redshift, the FWHM, the flux, and the luminosity for the emission line model, respectively. The values in parentheses are 1-$\sigma$ errors. Parameters fixed in the fitting with the code {\it pyspeckit} show errors with a value of 0.00 and a mark of $^*$.
    \end{tablenotes}

\end{table*}

\section{Properties}
\label{sec:properties}

\subsection{Galaxy Properties}
\label{sec:gal}
\subsubsection{Optical Photometric and Spectroscopic Properties}
\label{optical_pro}
As {\it Gaia} can easily detect close pairs (brighter than {\it G} $<$ 20.7) with its high spatial resolution of 0\farcs18~\citep{Gaia2016}, we first check the {\it Gaia} DR3 \citep{Gaia2022} detections of DPGPs.
There are 4 out of the 5 DPGPs with {\it Gaia} detections, and all of them are isolated detections, suggesting a single core profile in the center of the galaxy. However, it is still possible that these DPGPs include another faint core with {\it G} $>$ 20.7, which is below the current detection limit of {\it Gaia}. Another possibility is that the two cores are too close to be resolved by {\it Gaia}. 

\begin{figure*}
    \centering
    \includegraphics[width=0.15\textwidth]{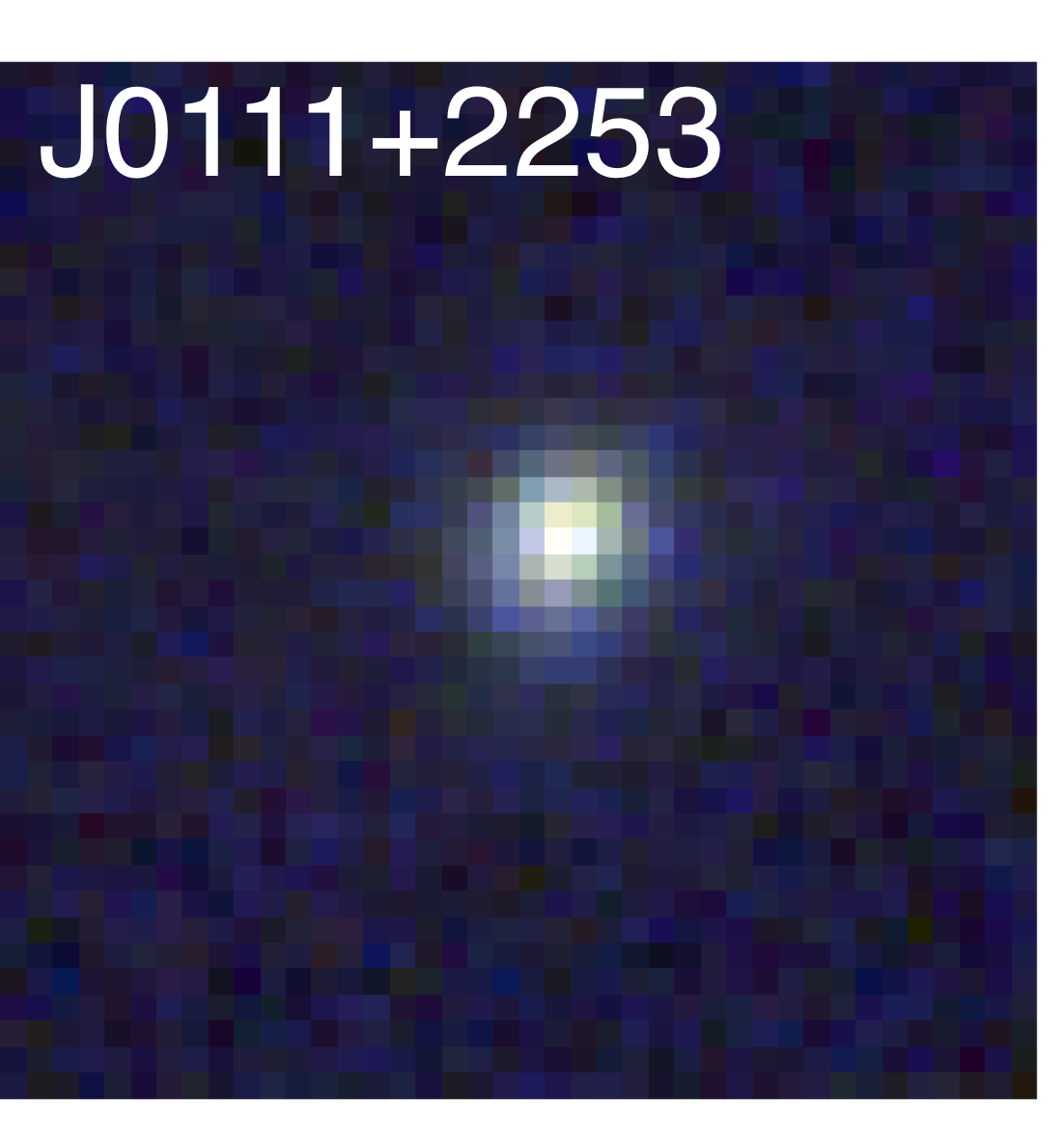}
    \includegraphics[width=0.8\textwidth]{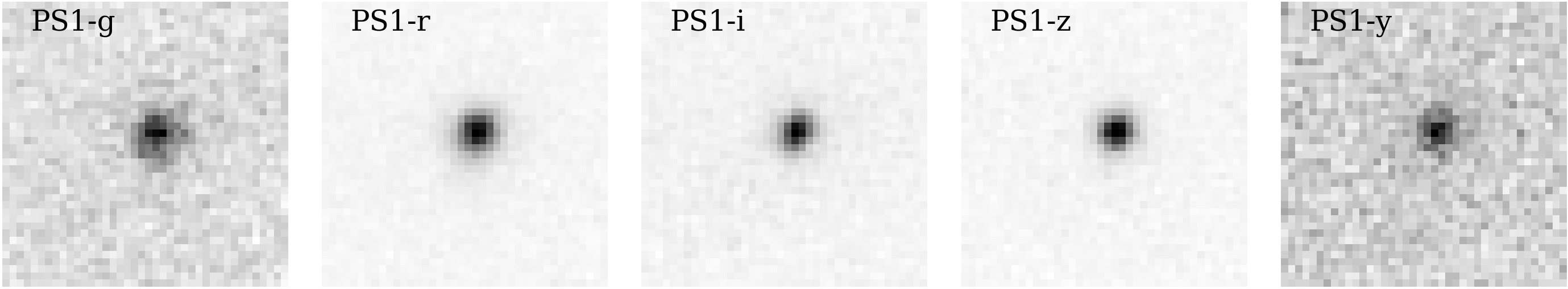}
    \includegraphics[width=0.15\textwidth]{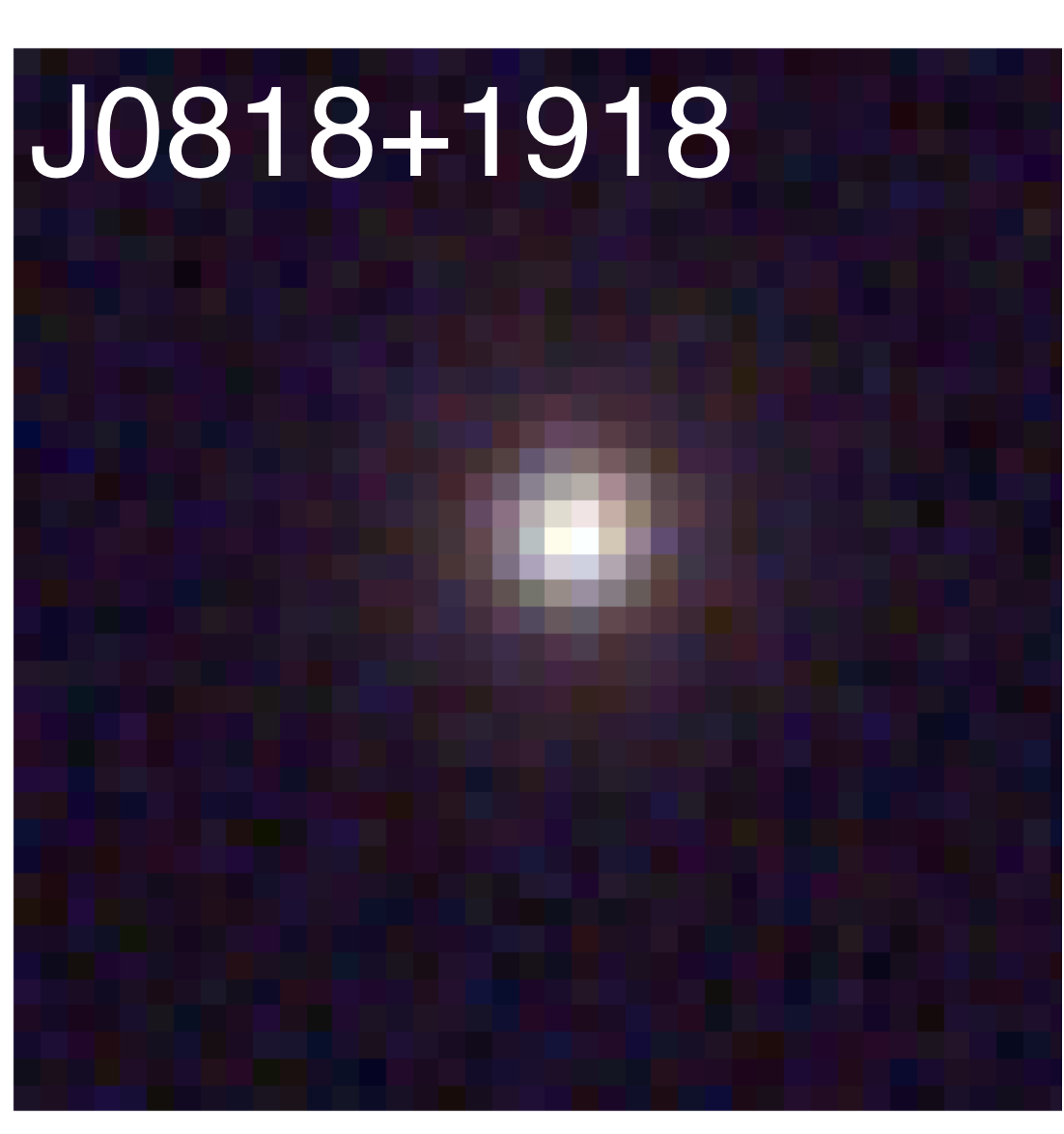}
    \includegraphics[width=0.8\textwidth]{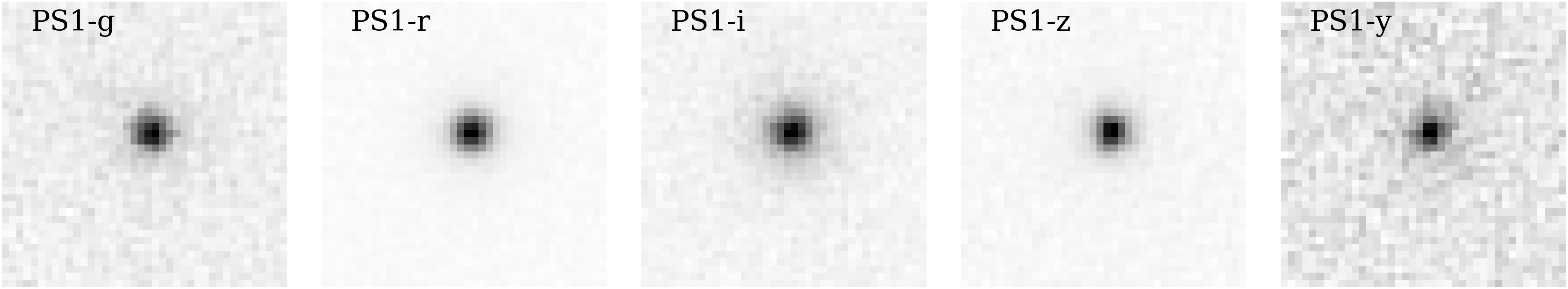}    \includegraphics[width=0.15\textwidth]{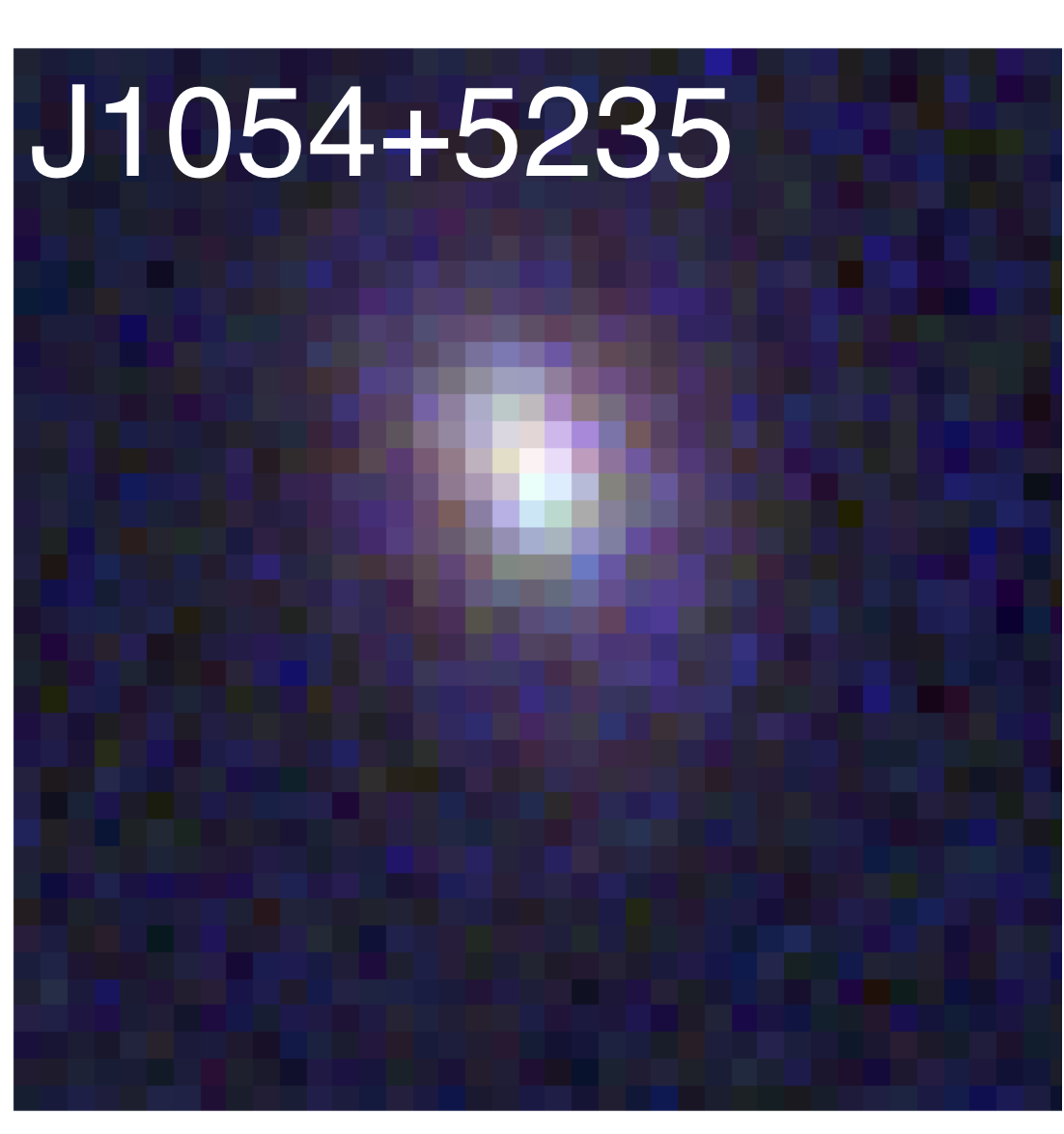}
    \includegraphics[width=0.8\textwidth]{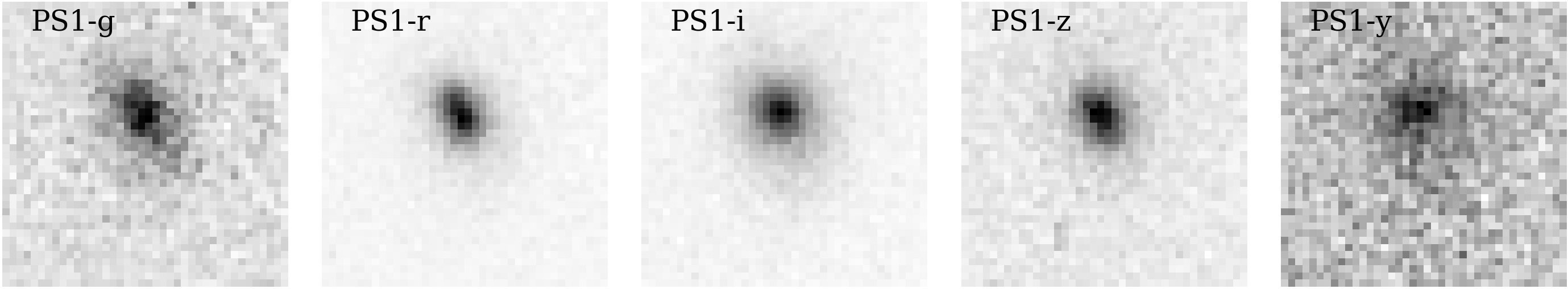}
    \includegraphics[width=0.15\textwidth]{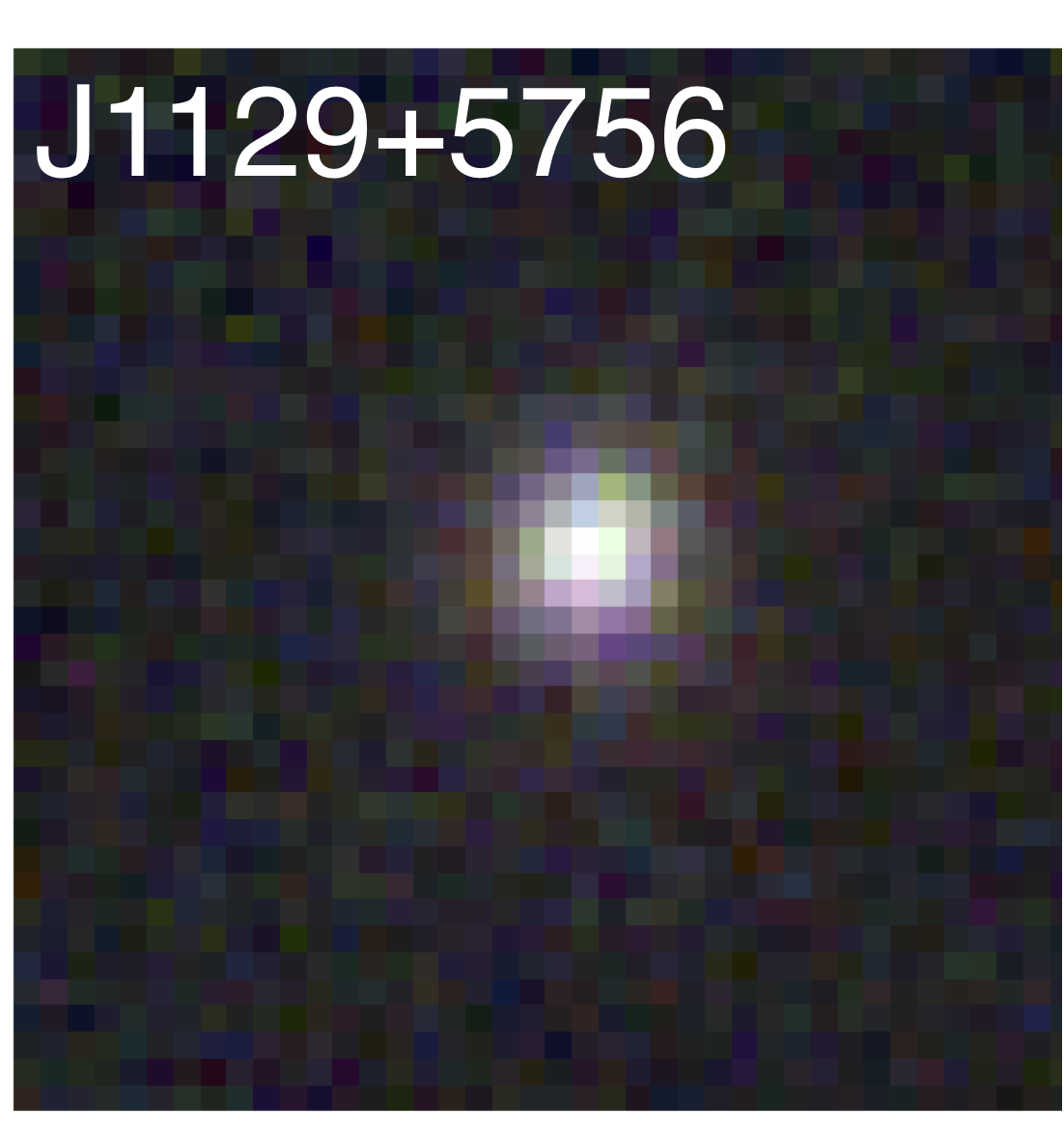}
    \includegraphics[width=0.8\textwidth]{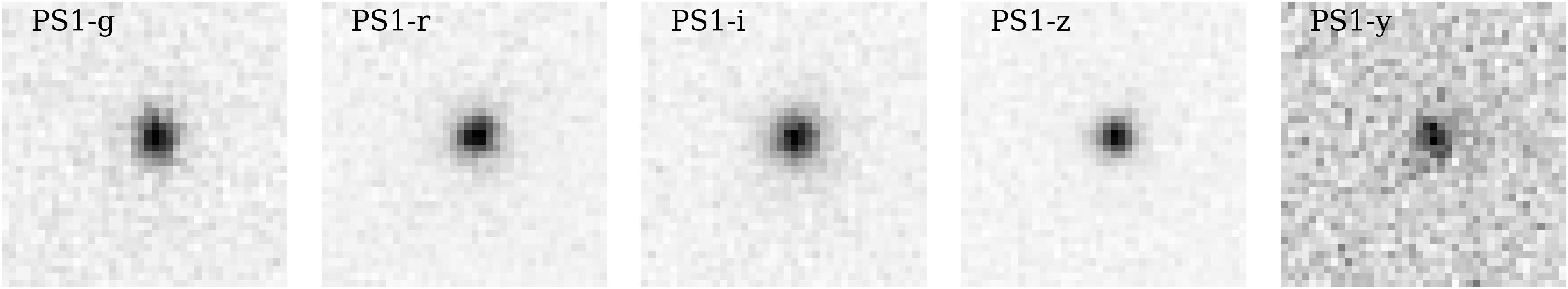}
    \includegraphics[width=0.15\textwidth]{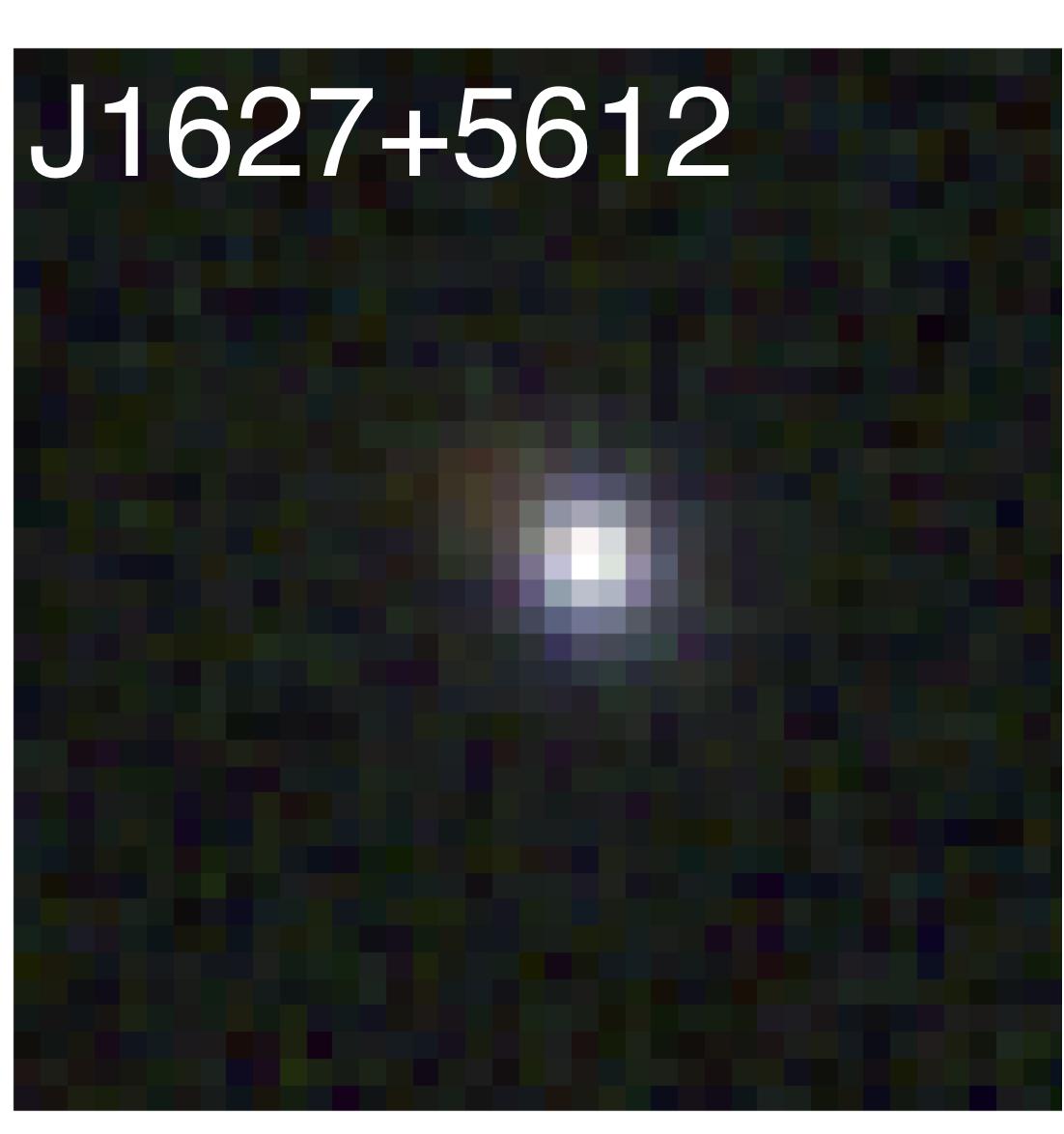}
    \includegraphics[width=0.8\textwidth]{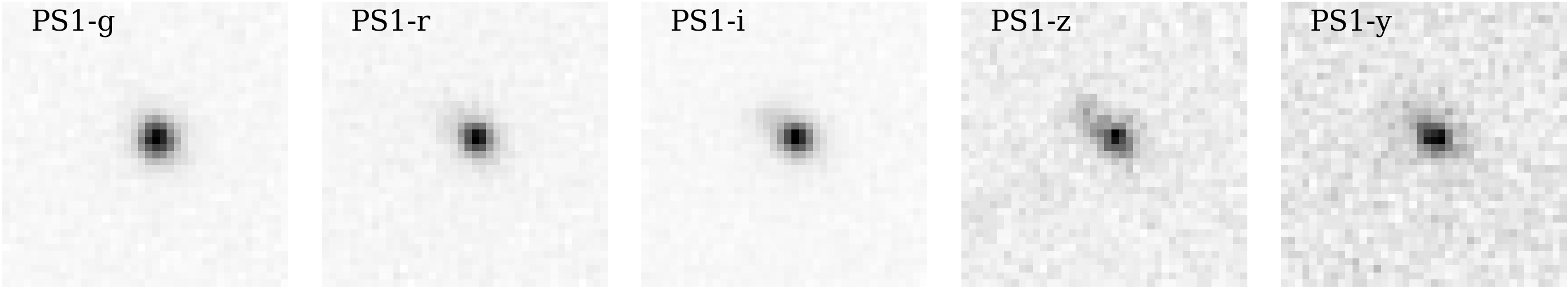}
    \caption{Pan-STARRS {\it gri}-color images (column 1) and {\it grizy} bands images (column 2-6) of the 5 DPGPs. The cutout size of each image stamp is 10$\arcsec\times$10$\arcsec$.}
    \label{fig:panstarrs}
\end{figure*}

We collect the optical imaging data from Data Release 17 (DR17~\footnote{https://www.sdss4.org/dr17/data\_access/}) of SDSS and the Data Release 1 and 2 \citep[DR1 and DR2 \footnote{https://outerspace.stsci.edu/display/PANSTARRS/}, ][]{Flewelling2020} of Panoramic Survey Telescope and Rapid Response System \citep[Pan-STARRS,][]{Chambers2016} surveys for our DPGPs. SDSS has median seeing values of 1\farcs52, 1\farcs44, 1\farcs32, 1\farcs26, 1\farcs29 and 5$\sigma$ depths of 22.15, 23.13, 22.70, 22.20, 20.71 mag for {\it ugriz} bands, respectively. 
The Pan-STARRS imaging data have slightly better spatial resolution with median seeing values of 1\farcs31, 1\farcs19, 1\farcs11, 1\farcs07, 1\farcs02 and higher sensitivity of 5$\sigma$ depths of 23.3, 23.2, 23.1, 22.3, 21.3 mag for {\it grizy} bands, respectively. Among this DPGP sample, two objects show slightly resolved structures in Pan-STARRS images (see Figure~\ref{fig:panstarrs}). J1627+5612 shows a fainter component in the northeast with an angular separation of $\sim$ 1\farcs3 from the brighter component. J1054+5235 has a slightly extended morphology along the northeast direction. 
These kind structures are also seen in HST images of some GP galaxies (see Figure 7 in \citealt{Cardamone2009} and Figure 15 in \citealt{Keel2022}).

The fitting results of double-peaked emission lines show that all of our DPGPs have more prominent blueshifted components than redshifted components. The \oiii\ velocity offsets $\rm V_{[OIII],off}$ between blueshifted and redshifted components range from 306 to 518 $\rm km\,s^{-1}$ (see Table~\ref{tab:property}). These velocity offsets are consistent with other double-peaked galaxies \citep[e.g., ][]{Comerford2012,Wang2019}. For either blueshifted or redshifted narrow components of \oiii$\lambda5007$ line of our DPGPs, the FWHMs range from 263 to 441 $\rm km\,s^{-1}$ (corresponding to the velocity dispersion in the range of 111 - 187 $\rm km\,s^{-1}$). Note that we cannot determine the systematic redshift of host galaxies because the stellar continuum is too weak (as mentioned in \S~\ref{sec:spectralAnalysis}), therefore we cannot compare the velocities of the two components to their host galaxies.

\begin{table*}
\centering
\caption{The galaxy properties and AGN selection of the 5 DPGPs.}
\label{tab:property}
\setlength{\tabcolsep}{1mm}{
\begin{tabular}{ccccccccccccc} 
\hline \hline
\multicolumn{1}{c}{\multirow{2}{*}{ID}} & \multicolumn{5}{c}{Galaxy properties} & & \multicolumn{4}{c}{AGN selection} \\
\cline{2-7} \cline{9-13}
 & \multirow{2}{*}{log(M$\rm _*/M_\odot$)} &
 \multirow{2}{*}{12+log(O/H)} & 
 SFR$_{\rm [OIII]}^a$ & 
 SFR$_{\rm H\alpha}^b$ & 
 SFR$_{\rm SED}^c$ & 
 SFR$_{\rm 1.4GHz}^d$ & & 
  \multirow{2}{*}{BLAGN?} & 
 \multirow{2}{*}{BPT-AGN?} & 
 \multirow{2}{*}{MIR-AGN?} & 
 \multirow{2}{*}{Radio-AGN?} & 
 \multirow{2}{*}{SED-AGN?} \\ 
 & & & M$_\odot\,yr^{-1}$ & M$_\odot\,yr^{-1}$ & M$_\odot\,yr^{-1}$& M$_\odot\,yr^{-1}$& & & & & & \\
 \hline

1 & 10.05 (0.15) & 8.74 (1.17) & 156.92 (1.62) & 39.17 (0.71)   & 25.42 (1.91) & -- & & FALSE &TRUE & TRUE & -- & TRUE \\
2 & 10.13 (0.05) & 8.78 (1.07) & 153.09 (3.41) & 53.59 (2.12) & 36.28 (1.93) & 208.71 (23.95)  & & TRUE & TRUE & TRUE & TRUE & FALSE \\  
3 & 10.72 (0.09) & 9.13 (0.90) & 265.32 (6.26) & 74.31 (2.98)  &  10.13 (0.82) & 549.76 (31.94)  & & TRUE &TRUE & TRUE & TRUE & TRUE \\ 
4 & 9.39 (0.04) & 8.77 (1.12) & 104.95 (2.57) & 29.80 (1.94) & 54.98 (2.75)  & 416.29 (25.94) & & FALSE &TRUE & TRUE & TRUE & TRUE \\
5 & 9.94 (0.16)  & 8.50 (0.01) & 42.08 (2.29) & -- & 21.80 (10.04) & -- & & -- &-- & TRUE & -- & TRUE \\

\hline
\end{tabular}
}
    \begin{tablenotes}
    \item NOTE. $^a$ The SFR estimated by using \oiii\ emission line with the relation from \cite{Villa2021}. $^b$ The SFR estimated by using \ha\ emission line with the relation from \cite{Kennicutt2012}. $^c$ The SFR derived from the SED fitting. $^d$ The SFR derived from the VLA-FRIST 1.4GHz flux following \citet{Murphy2011}. The values in parentheses are 1-$\sigma$ errors.
    \end{tablenotes}
\end{table*} 

\subsubsection{Stellar Mass}
To obtain the stellar masses of DPGPs, we model the SED using CIGALE \citep{Burgarella2005,Noll2009,Boquien2019} with available ultraviolet (UV), optical, and mid-infrared data observed by GALEX \citep{Morrissey2007}, SDSS, and WISE \citep{Wright2010}, respectively. We use a series of synthesis models of delayed star-forming history, single stellar population \citep{BC03}, IMF \citep{Chabrier2003}, nebular lines \citep{Inoue2011}, dust emission \citep{Dale2001,Dale2014}, attenuation \citep{Charlot2000}, and AGN \citep{Fritz2000,Stalevski2016} models.

\begin{figure}
    \centering
    \includegraphics[width=0.4\textwidth]{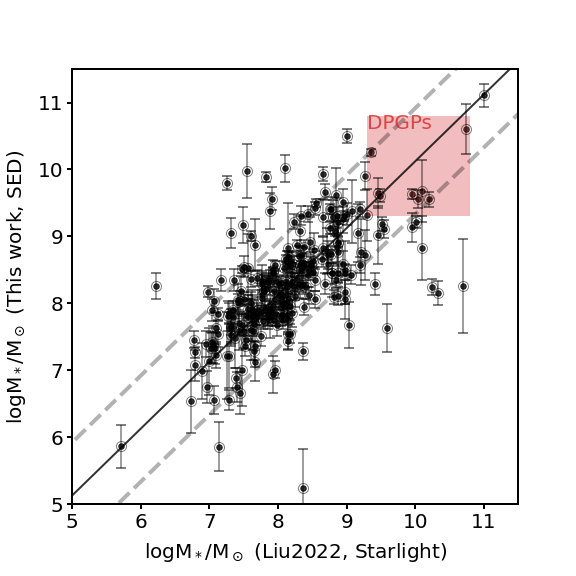}
    \caption{
    Comparison of stellar mass between this work and \citet{Liu2022}. The solid line represents the liner fitting and the dashed lines represent the $\pm$1$\sigma$ scatters. Since none of our DPGPs are included in 259 galaxies with stellar masses calculated by \citet{Liu2022}, we mark a red shaded region to show the range of stellar masses of our DPGPs.
    }
    \label{fig:comparisonMass}
\end{figure}

\cite{Liu2022} derived stellar masses for 259 GPs using {\it STARLIGHT} \citep{Cid_Fernandes2011} by directly modeling LAMOST spectra. However, none of the DPGPs is included. Thus, we first examine the quality of stellar mass derived from the SED fitting by comparing our result with that of \cite{Liu2022}. For this sample of 259 GPs with known stellar mass, our results via the SED fitting are slightly higher than that of  {\it STARLIGHT} and the difference between these two methods is $\rm log(M_{*,Starlight}-M_{*,SED}) = -0.13\pm0.54$ (see Figure~\ref{fig:comparisonMass}). 

We then derive the stellar masses of DPGPs by SED fitting and also apply this method to other DPGP candidates for comparison. Our DPGP candidates have the stellar mass in the range of 10$^{9.4}-10^{10.7}$$\ M_\odot$,  which are $\sim$ 20-400 times the median mass (10$^{8.12}\ M_\odot$) of the 259 GP sample. Our DPGPs show higher stellar masses compared to DPGP candidates and the other GPs of the parent sample (see Table~\ref{tab:property} for the specific value of mass). However, our DPGPs have slightly lower mass compared with other double-peaked narrow emission line galaxies, which have larger mass of $\rm M_* > 10^{10.4}$ \citep{Liu2010,Maschmann2020}.

\subsubsection{Metallicity}
\label{sec:metallicity}
We estimate the gas-phase metallicity of all DPGP candidates using the R23 (R23 = (\oii $\lambda$3727 + \oiii $\lambda\lambda$4959,5007) / \hb) method following \cite{Tremonti2004} for low redshift star-forming galaxies (SFGs). Since this candidate DPGP sample has broader line width, which tends to host AGN, the empirical relation of \cite{Tremonti2004} would underestimate the metallicity of AGN in our sample. Thus, we identify the BPT-selected AGN in the candidate DPGP sample as described in \S~\ref{sec:agn}, and then use the empirical relation of \cite{Dors2021} for AGN. Therefore, we obtain an average oxygen abundance of 12+log(O/H) = 8.78 for DPGPs (see Table~\ref{tab:property} for details) and 8.65 for DPGP candidates. This result is $\sim$ 0.5 dex higher than the median value of 12+log(O/H) from \cite{Liu2022} using N2 method for the star-forming galaxies, suggesting that stronger evolution in DPGPs or DPGP candidates than other star-forming galaxies, which may be related to the interaction between AGN and the environment.

\begin{figure}
    \centering
    \includegraphics[width=0.4\textwidth]{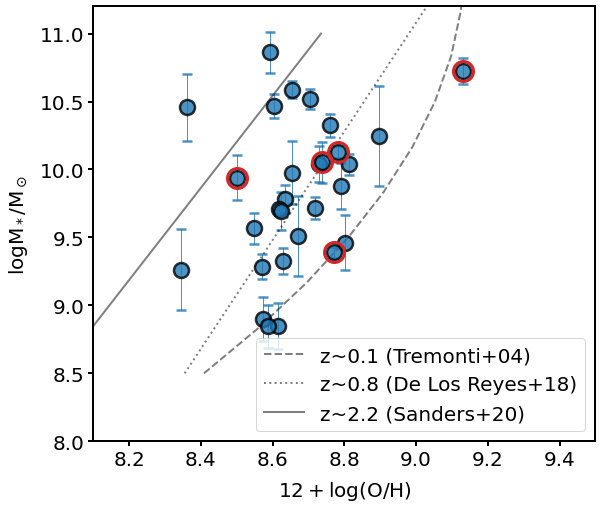}
    \caption{Stellar mass and gas-phase metallicity of the DPGP candidates (blue filled circles) and the 5 DPGPs (red open circles). The lines represent the mass-metallicity relation of SFG at z $\sim$ 0.1 \citep{Tremonti2004}, at z $\sim$ 0.8 \citep{de_los_Reyes2015}, and z $\sim$ 2.2 \citep{Sanders2020}, respectively.}
    \label{fig:mass_metallicity}
\end{figure}

We present the diagram of stellar mass versus metallicity for our DPGPs in Figure~\ref{fig:mass_metallicity}. The mass-metallicity relation of DPGPs locates between that of SFG at z $\sim$ 0.1 \citep{Tremonti2004}, at z $\sim$ 0.8 \citep{de_los_Reyes2015}, and at z $\sim$ 2.2 \citep{Sanders2020}. Overall, these DPGPs have large scatter in stellar mass-metallicity relation. 

\subsubsection{SFR}
The H$\alpha$ emission line is a direct indicator of star-formation rate (SFR) of galaxies~\citep{Kennicutt1998,Kennicutt2012}, which could be described as:
\begin{equation}
  {\rm SFR}\,({ M_\odot\,\rm yr^{-1}})\, =  5.3 \times 10^{-42}\, L_{\rm H \alpha} \,{\rm(erg\,s^{-1}}).
\end{equation}
Note that we correct \ha\ line luminosity for the dust attenuation here, and the total \ha\ line flux is used. 

 However, \ha\ lines could only be well modeled for 4 of 5  DPGPs (as mentioned in \S \ref{sec:spectralAnalysis}).
Because of the proportional relation between the strength of the \ha\ line and other emission lines such as \oiii\ and \oii\ lines \citep[and references therein]{Suzuki2016}, the \oiii\ emission line could be used to estimate SFR \citep{Hippelein2003, Ly2007, Straughn2009}. 
We thus calculate the SFRs of DPGPs using the total \oiii$\lambda5007$ emission lines, following the relation in \cite{Villa2021} which corrected the dust attenuation of \oiii\ through SED fitting and had the relation between the SFR and \oiii\ luminosity as:
\begin{equation}
  {\rm SFR}\,(M_\odot\,{\rm yr^{-1})}= 6.3 \times 10^{-42}\,L_{[{\rm O\sc III}]}\,\rm{(erg\,s^{-1})}.
\end{equation}

We also drive SFRs using the radio emission (see \S\ref{sec:vla} in detail) and compile the SFRs derived from the SED fitting.
The various SFRs of these DPGPs are listed in Table~\ref{tab:property}.
These DPGPs show intense star formation rates from \ha\ or \oiii\ emission lines of SFR $>29\, M_\odot\,yr^{-1}$. However, these sources all exhibit significant AGN activity (see \S\ref{sec:agn} for details), and therefore AGN make a non-negligible contribution to their nebular emission lines \citep{Kauffmann2009,Jin2021}.  All DPGPs exhibit excesses on SFRs derived from the \oiii\ and 1.4 GHz luminosity when compared to the SFR derived from the SED fitting, which further suggests the presence of AGN in these galaxies. The SFRs derived from \ha\ luminosity are similar to those derived from SED fitting, although the values are not exactly same due to the effects of continuum subtraction in the spectral fitting and the AGN contribution in SED fitting.

\subsection{AGN Activities}
\label{sec:agn}
We identify the AGN in DPGPs by applying five methods which include \ha\ line width, emission-line diagnostic \citep[BPT diagram,][]{Baldwin1981,Kewley2006}, mid-infrared color, radio radiation, and AGN fraction derived from SED fitting.  The broad-line region of AGN has a velocity of $\rm FWHM_{H\alpha}\, \gtrsim\, 1000 \, km\,s^{-1}$. Emission-line diagnostic can effectively classify the star-forming galaxies and AGN by the hardness of radiation. Mid-infrared color presents the radiation of the dust torus of AGN. Radio emission directly traces the radio jet, which is also powered by the accretion system around the central massive black hole. AGN fraction from SED fitting presents the AGN contribution to the continuum emission of the host galaxy. 

\subsubsection{Line Widths}
For this DPGP sample, four galaxies (J0111+2253, J0818+1918, J1054+5235, and J1129+5756) have measurements of \ha\ lines. Based on the fact that AGN have typical line widths of FWHM$\rm_{H\alpha}\, \gtrsim\, 1000\, km\,s^{-1}$ \citep{Weedman1977}, we classify 2 (J0818+1918 and J1054+5235) out of these 4 DPGPs as AGN (see Table \ref{tab:fittingResult} in details). J1627+5612 has no coverage of the \ha\ and \nii\ lines in its spectrum.

\subsubsection{Emission-line Diagnostic}
\label{sec:bpt}

\begin{figure}
    \centering
    \includegraphics[width=0.45\textwidth]{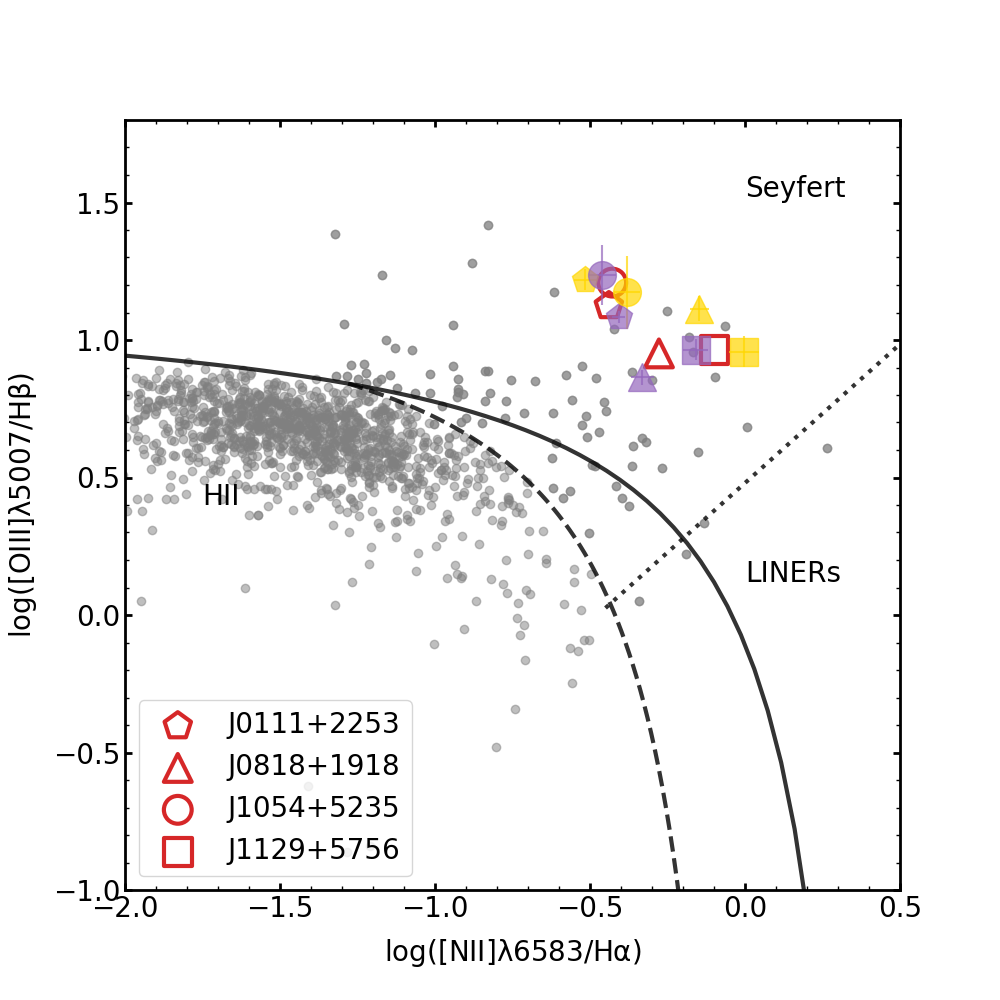}
    \caption{BPT diagram for \greenpea\ galaxies. The gray dots represent GPs measured by \citet{Liu2022}. The DPGPs whose \ha\ complex have been well modeled are highlighted. The diamond, triangle, circle, and square symbols represent DPGP J0111+2253, J0818+1918, J1054+5235, and J1129+5756, respectively. The integral spectra, redshifted components, and blueshifted components of the narrow emission lines are marked in red, yellow, and purple, respectively. The classification lines follow \citet{Kewley2001}(solid line) and \citet{Kauffmann2003}(dashed and dotted lines).}
    \label{fig:bpt}
\end{figure}

With the classification criteria of \cite{Kewley2001} and \cite{Kauffmann2003} \citep[also see][]{Kewley2006}, those four DPGPs having well modeled \ha\ and \nii\ lines are identified as AGN in the BPT diagram (see Figure~\ref{fig:bpt}). Furthermore, their blueshifted and redshifted components are also classified as AGN, inferring the existence of dual AGN. 
J1627+5612 has a $\rm log([OIII]\lambda5007/H\beta)$ = 0.6 and 0.6 for the redshifted and blueshifted components, while its spectrum doesn't cover the \ha\ and \nii\ lines, which prevent us to judge if it is an AGN from the BPT diagram.
Overall, there are 4 out of 5 DPGPs are selected as AGN via the narrow emission-line diagnostic. 

\subsubsection{Mid-infrared Observation}
\label{sec:MIR}

\begin{figure}
    \centering
    \includegraphics[width=0.4\textwidth]{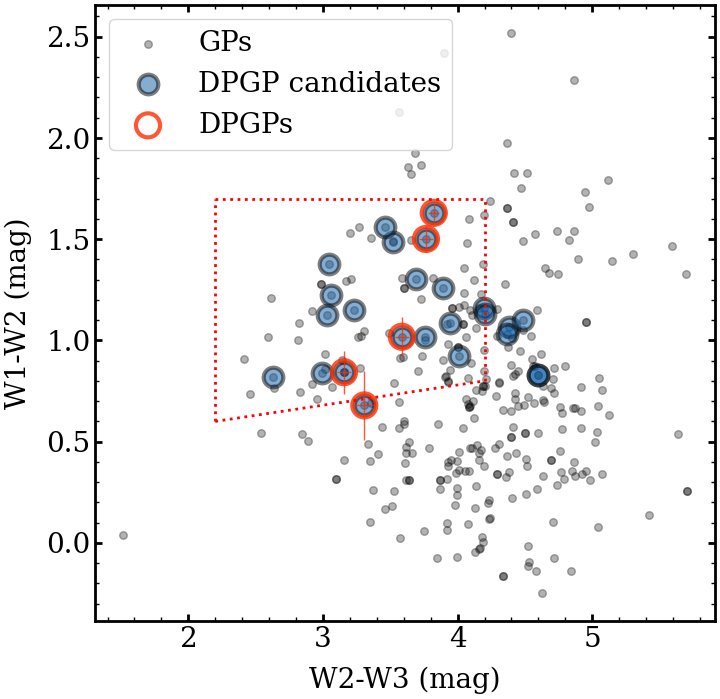}
    \caption{WISE color-color diagram for \greenpea\ galaxies.  DPGPs with WISE detection are marked in red. The sample of DPGP candidates, mentioned in \S \ref{DPGP_can}, are also illustrated as blue circles.
    The red dashed lines mark the mid-infrared AGN selection criteria  from \citet{Jarrett2011}.
    }
    \label{fig:mid_agn}
\end{figure}

We collect the mid-infrared information of DPGPs from the ALLWISE source catalog\footnote{https://wise2.ipac.caltech.edu/docs/release/allwise/}. 
All DPGPs show mid-infrared detections with S/N in a range of $\sim$ 26-36, 17-40, and 7-31 in W1, W2, and W3 bands, respectively. To classify the possible origin of the strong mid-infrared radiation of DPGPs, we follow color-color criteria in \cite{Jarrett2011} and present the WISE color-color distribution in Figure~\ref{fig:mid_agn}. There are 4 out of 5 DPGPs located in the mid-infrared AGN region. For the remaining one, it could also be identified as mid-infrared AGN when considering its photometric uncertainties in the WISE color.
Therefore, the mid-infrared radiation of all these 5 DPGPs
would be mainly contributed by AGN activities.


\subsubsection{Radio Observation}
\label{sec:vla}
The Faint Images of the Radio Sky at Twenty-cm \citep[FIRST~\footnote{http://first.astro.columbia.edu},][]{Becker1995} survey had used the NRAO Very Large Array (VLA) to explore the radio radiation of optical sample at $\sim$ 1.4 GHz, covering over 10000 deg$^2$ and reaching a depth of 0.15 mJy. We explore the radio properties of this DPGP sample with FIRST. There are 3 DPGPs, J1054+5235, J1129+5756, and J0818+1918, with radio detections of 
$f_{\rm 1.4GHz}$ = 2.58, 2.07, and 0.95 $mJy$, corresponding to SFR$_\textrm{radio} = $ 549, 416, and 209  $M_\odot\,yr^{-1}$ \citep[with the conversion from][]{Murphy2011}, respectively. These DPGPs show large SFRs which significantly exceed the SFRs derived from \ha\ luminosity and SED fitting (see Table \ref{tab:property}), indicating the excess contributed by AGN. The results here are consistent with the results from
the emission-line diagnostic (\S~\ref{sec:bpt}) and the mid-infrared observation (\S~\ref{sec:MIR}).

\begin{figure}
    \centering
    \includegraphics[width=0.25\textwidth]{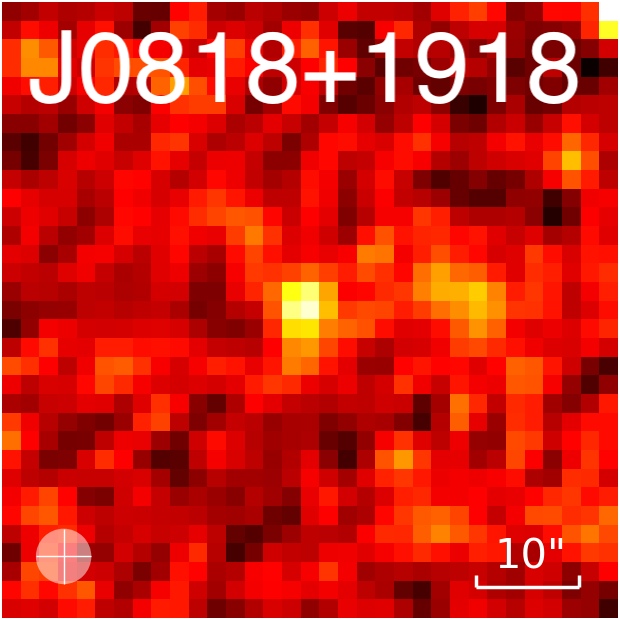}
    \includegraphics[width=0.25\textwidth]{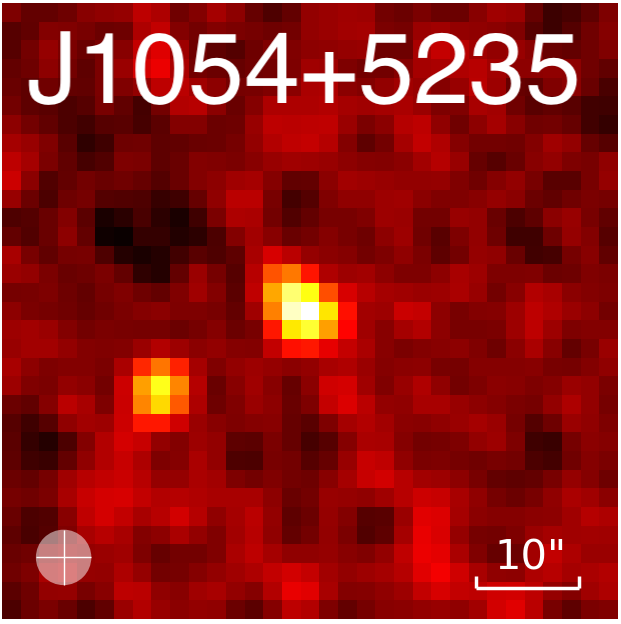}
    \includegraphics[width=0.25\textwidth]{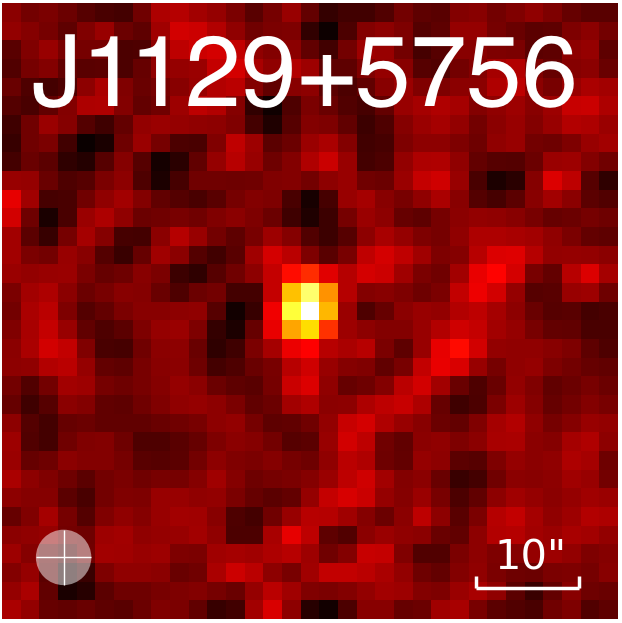}
    \caption{VLA-FIRST 1.4 GHz imaging cutouts of  J0818+1918, J1054+5235, and J1129+5756 in a size of 1\arcmin$\times$1\arcmin. The FIRST images have a beam size of 5\farcs4. All the three DPGPs are unresolved with the VLA-FIRST survey.
    }
    \label{fig:vla}
\end{figure}

We also present the FIRST images of these three DPGPs in Figure~\ref{fig:vla}. 
However, with the FIRST's resolution of $\sim$ 5\arcsec, we are unable to distinguish further structures of these objects from the current radio images.

\subsubsection{AGN Fraction in SED Fitting}
We use the AGN model from \citet{Stalevski2016} in the SED fitting to probe the AGN contribution in the infrared emission. The AGN fraction $f_{\rm AGN}$ in CIGALE is defined as the ratio of the AGN luminosity to the sum of the AGN and dust luminosities. We set the input AGN fraction in the range of 0-0.9 with a step of 0.1, and use a cut of 0.2 \citep[as used in][]{Liu2022} to select AGN from SED fitting. Most (4/5) of DPGPs have the AGN fraction of $f_{\rm AGN} > 0.2$, implying the substantial contribution of AGN activities.

\subsubsection{Final Classification}
Combining the five methods above, 4 DPGPs can be classified as AGN by at least three methods (see Table~\ref{tab:basicProperty} for details). The remaining one, J1627+5612,  can be identified as AGN via the mid-infrared method and the SED fitting method. Therefore, all of these DPGPs are reliable AGN.  

\section{Discussion}
\label{sec:discussion}
\subsection{Nature of DP}
\subsubsection{Are DPGPs the NLS1 Galaxies?}
\label{sec:nls1}
Narrow-Line Seyfert 1 (NLS1) galaxies are regarded as evolutionarily young objects compared to the broad-line galaxies \citep{Ryan2007}. They are characterized by intermediate line widths as $\rm 500<FWHM_{H\beta}<2000\, km\, s^{-1}$, which are broader than type 2 AGN but narrower than typical broad-line Seyfert galaxies \citep{Osterbrock1985}. NLS1 galaxies also have strong Fe{\sc ii} and relatively week \oiii\ lines \citep{Goodrich1989} with \oiii$\lambda5007\AA$/\hb\ $\leq$ 3. The \oiii\ profiles of NLS1s often show blue wings, indicating outflows and winds triggered in the nuclear narrow-line region \citep{Marziani2003, Schmidt2018}.

Our DPGPs have similar \hb\ line widths as NLS1 galaxies. However, DPGPs differ from the NLS1 galaxies for other line properties. As presented in Table \ref{tab:fittingResult}, all of our DPGP sample have extremely strong \oiii\ intensity with \oiii$\lambda5007\AA$/\hb\ $\gg$ 3, unsatisfying the common criteria for NLS1 galaxies \citep[e.g.,][]{Osterbrock1985,Goodrich1989,Williams2002,Schmidt2016,Schmidt2018}. More often, the blue wings of \oiii\ lines in NLS1 galaxies are weaker compared to the core components. In contrast, for DPGPs the blueshifted components are more luminous and have comparably narrow line widths to the redshifted components. 

Overall, the DPGPs and NLS1 show similar line widths of \hb\ lines, but the \oiii/\hb\ ratio and line profiles are significantly different, suggesting that the nature of the DPGP differs from that of NLS1.

\subsubsection{Are DPGPs the Type 2 Quasars?}
Type 2 quasars are defined as a kind of galaxies with narrow \hb\ FWHM, high ionization condition, and extreme high narrow-line \oiii\ luminosity as type 1 quasars \citep{Zakamska2003,Reyes2008}. These selection criteria could be described as:
\begin{itemize}
    \item[(a)] $\rm FWHM_{H\beta} < 2000\, km\,s^{-1}$,
    \item[(b)] Be classified as AGN in the BPT diagram or ($\rm log([OIII]\lambda5007/H\beta) > 0.3\ \&\ FWHM_{[OIII]} > 400\, km\,s^{-1}$) , and
    \item[(c)] $\rm L_{[OIII]} > 10^{41.9} erg\,s^{-1}$.
\end{itemize}
All of the DPGPs meet with the criterion (a) as described in \S\ref{sec:nls1}. For the criterion (b), 4 out of 5 DPGPs meet with the former criterion (as described in \S\ref{sec:bpt}), and the remaining one, whose \ha\ is not covered by its spectroscopic range, satisfies the latter criterion (see \S\ref{sec:bpt} and Figure \ref{fig:LvsFWHM}). Our DPGPs also meet the criterion (c), as they all have the $\rm L_{[OIII]} >10^{42} erg\,s^{-1}$ (see Table \ref{tab:fittingResult} and Figure \ref{fig:LvsFWHM} in details). Therefore, all of our DPGPs meet the selection criteria for type 2 quasars.

Interestingly, J1129+5756 has been reported in \cite{Villar2012} as a type 2 quasar. The HST/WFPC2 F814W image (HST proposal ID: 10880) of J1129+5756 shows an irregular structure, inferring the  merger/interaction processing.

\subsubsection{What Drives the Double-peaked \oiii\ Lines in DPGPs?}
The double-peaked profiles of narrow emission lines are possibly originated from three hypothesis, the outflow, the rotating disk, or dual AGN. Below we discuss the likelihood of each origin. 
\begin{itemize}
    \item {\it AGN-driven  outflow.} In the simulation of AGN-driven outflow, \cite{Bae2016} constrained the intrinsic velocity of outflow in the range of 500 - 1000 $\rm km\,s^{-1}$ for most AGN. In observations, several authors demonstrated that the outflow has a velocity dispersion of $\sigma > 400\, \rm km\,s^{-1}$ \citep{Muller-Sanchez2015, Woo2016} and results in a line-of-sight velocity of \oiii\ line of $\rm V > 400\, km\,s^{-1}$ \citep{Das2006, Fischer2013, Crenshaw2015} which can generate the double-peaked profile for a spectral resolution of LAMOST and SDSS. However, our DPGPs have low velocity dispersion ($\sigma < 200 \rm \, km\,s^{-1}$, as mentioned in \S\ref{optical_pro}). We note that we do not compare the velocity of each component of our DPGPs with the outflow, because the velocities of host galaxies could not be determined.
    In addition, the outflow often causes a profile of core components along with blue wings \citep{Karouzos2016}. The red wings are usually weak and even cannot be seen, because of the obscuration by the galaxy disks or AGN dust torus \citep{Bae2016}. However, all DPGPs show more prominent and narrow blueshifted components.
    We regard that the nature of double-peaked narrow-line profile of our DPGP sample is less likely to be outflows. 
    Despite AGN-driven outflow may not contribute to the observed double-peaked narrow emission line, they could still exist as manifested by the broad \oiii\ component (such as, J0818+1918, J1054+5235, and J1129+5756). Compact radio jets could induce turbulence in the ambient gas and result in large \oiii\ line width as this broad \oiii\ emission is only visible in radio-AGN \citep{Holt2006,Holt2008,Mullaney2013}.
    
    \item {\it Rotating disk of host galaxies.} Ionized gas rotating in galaxy disks with different rotating velocities could produce the double-peaked profiles of emission lines \citep[]{Elitzur2012,Kohandel2019}. 
    This profile is characterized by a low velocity of $\rm |V| < 400\, km\,s^{-1}$ and a low velocity dispersion of $\sigma < 500\, \rm km\,s^{-1}$ \citep{Smith2012}, which is consistent with our DPGP sample. 
    The double-peaked profile caused by the rotating disk appears highly symmetrical. However, J0111+2253 and J1627+5612 are unlikely to fit this scenario, as their line profiles show asymmetry (the blueshifted components are at least 1.5 times brighter than the redshifted components at peak). J0818+1918, J1054+5235, and J1129+5756 have symmetric profiles for emission lines, while J0818+1918 and J1129+5756 do not appear obviously inclined and therefore have a less possibility to identify a double-peak feature caused by the rotating disk \citep[see Fig. 4 in][]{Maschmann2023} except for a high speed disk case.  The rotating disk could explain the origin of double-peaked narrow lines of J1054+5235. Overall, the rotating disks are unlikely to be the origin of the double-peaked profile for most of our DPGP sample.
    
    \item {\it Dual AGN.} The rotation of pc/kpc-scale dual AGN could explain the double-peaked profiles. As described in \S \ref{sec:bpt}, most of our DPGPs(J0111+2253, J0818+1918, J1054+5235, and J1129+5756) exhibit two AGN components in the BPT diagram (see Figure \ref{fig:bpt}), suggesting that they are more likely to be dual AGN. Given the $\rm M_{BH}-\sigma_{[OIII]}$ relation derived by \cite{Bennert2018}, we could estimate the mass of each black hole in a dual AGN directly by using the velocity dispersion of \oiii\ narrow line. With an assumption that the double-peak profiles of our DPGPs originate from dual AGN, each DPGP hence have two black holes with masses in the range of $10^{7.2}-10^{8.0}\ M_\odot$ and $10^{7.2}-10^{8.1}\ M_\odot$ estimated from the blueshifted and the redshifted components, respectively.
    Overall, dual AGN is a possible origin of our DPGPs, yet observation with higher spatial resolution is needed for confirming dual AGN in these DPGPs.
\end{itemize}
 
Overall, given the current data, the double-peaked profiles of these five GPs are more likely originated from dual AGN rather than rotating disks or outflows. Yet, imaging observation with higher spatial resolution and spectroscopic observation with spatial information (i.e., long-slit spectrum and integral field spectrum) are needed to reveal the nature of these DPGPs.

\subsection{Comparison with Other Double-peaked \oiii\ Samples}
\label{sec:compare_DP}
Based on the database of the MPA-JHU SDSS DR7 galaxy catalog, \cite{Ge2012} had automatically selected a sample of double-peaked galaxies and asymmetric emission-line galaxies, composed of 3030 double-peaked emission-line galaxies. With the emission-line diagnostics of separated components, they had classified these objects into different types, such as the AGN-AGN pairs, AGN-galaxy pairs, and galaxy-galaxy pairs. The fraction of galaxy-galaxy pairs and AGN pairs (including AGN-AGN and AGN-galaxy pair) are 1\% and 1.5\% to the parent sample of ELGs, respectively. 

To explore the presence of \greenpea s in this sample of double-peaked narrow-line galaxies, we apply the photometric and spectroscopic methods of \cite{Cardamone2009} and \cite{Izotov2011} to filter the sample. We directly use the measurements of emission lines derived by the MPA-JHU team. \cite{Cardamone2009} selected GPs through the SDSS {\it gri} color and objects' size, therefore they selected the luminous compact objects with a redshift range of 0.112-0.360. In contrast, \cite{Izotov2011} directly searched from the SDSS spectroscopic survey, and thus selected luminous compact objects with strong \oiii\ lines at various redshifts. There are less than 0.1\% (2/3030) of galaxies that meet the criteria of \cite{Izotov2011}, while none satisfies the conditions of \cite{Cardamone2009}, because these criteria are more strict in redshift. 
We note that there are 837 type 2 AGN pair in this DP sample, however, none of them is identified as a GP, inferring that DPGPs differs from type 2 dual AGN in colors or \oiii\ equivalent widths.
Overall, \greenpea\ galaxies are rare in the sample of double-peaked emission-line galaxies.

The two double-peaked strong emission-line galaxies (J161555.1+420624.6 and J165844.41+351923.2) in \cite{Ge2012} are both galaxy-galaxy pairs and have redshifts of $z < 0.1$, of which the \oiii\ lines fall in the {\it g} band and show blue colors in SDSS images. They have similar spectral properties as our DPGPs, i.e., have double narrow emission-line profiles. Interestingly, one of these two galaxies has two detections with Gaia separated by 1.1\arcsec, suggesting that the double-peaked line profiles may originate from the ongoing merger of galaxies.

\section{Conclusions}
From the available \greenpea\ catalogs derived from LAMOST and SDSS surveys, we newly discover a sample of 5 double-peaked \greenpea s by analyzing the emission lines of their optical spectra. We perform a detailed modeling of the double-peaked line profiles, and explore galaxy properties and AGN activities of these DPGPs.
Our main results are summarized as follows:
\begin{itemize}
    \item These DPGPs can be well modeled using two narrow components or two narrow components plus a broad component. They have similarity of more prominent blueshifted components with \oiii\ luminosity of $> 10^{42}\, \rm erg\, s^{-1}\, cm^{-2}$. The velocity offsets between the blueshifted and redshifted components range from 306 to 518 $\rm km\, s^{-1}$. The line widths of individual \oiii$\lambda 5007 \AA$ narrow components are up to 441 $\rm km\, s^{-1}$.
    \item Using the SED fitting with photometric data from UV to mid-infrared, we estimate the stellar masses of $10^{9.39-10.72}\ \rm M_\odot$ for these DPGPs. Given the measurement of the emission-line ratio, we derive the gas-phase metallicities of 8.50 < 12+log(O/H) < 9.13. These DPGPs have larger masses and metallicities compared with other GPs. 
    \item With the \ha\ line width, emission-line diagnostic, mid-infrared color, radio radiation, and AGN fraction derived from SED fitting, we find that these DPGPs host AGN. 
    \item These DPGPs meet with the criteria of type 2  quasars.
    \item The BPT classification of two narrow components of 4 out of 5 DPGPs (i.e., J0111+2253, J0818+1918, J1054+5235, and J1129+5756) show evidences of dual AGN in the galaxy centers. However, we cannot exclude the other possibility (e.g., rotating disks or outflows) for the remaining one DPGP.
\end{itemize}
Overall, this DPGP sample is rare in both the GP galaxy sample and the double-peaked emission-line galaxy sample in the local universe. Although these objects are similar to the DP sample found in type 2 AGN \citep[e.g.,][]{Comerford2009,Wang2009,Ge2012,Muller-Sanchez2015,Liu2018}, they differ significantly in terms of colors or \oiii\ equivalent widths (see \S\ref{sec:compare_DP}). In galaxies
as compact as GPs, the mechanism  triggering such high \oiii\ luminosity as well as double-peaked emission lines in a compact volume remains an intriguing puzzle. The physical properties of these DPGPs provide a new chance to explore the growth of this kind of extreme sample and probe the analogs in the early universe. However, due to the small sample size, more reliable conclusions would be made with further analysis of the additional data.

\section*{Acknowledgements}
We would like to thank the anonymous referee for very helpful comments. Z.Y.Z. acknowledges the support by the National Science Foundation of China (12022303) and the China-Chile Joint Research Fund (CCJRF No. 1906).
We also acknowledge the science research grants from the China Manned Space Project with NO. CMS-CSST-2021-A04, CMS-CSST-2021-A07. F.T.Y. acknowledges the support by the Funds for Key Programs of Shanghai Astronomical Observatory (No. E195121009) and the Natural Science Foundation of Shanghai (Project Number: 21ZR1474300)

Guoshoujing Telescope (the Large Sky Area Multi-Object Fiber Spectroscopic Telescope LAMOST) is a National Major Scientific Project built by the Chinese Academy of Sciences. Funding for the project has been provided by the National Development and Reform Commission. LAMOST is operated and managed by the National Astronomical Observatories, Chinese Academy of Sciences.

Funding for the Sloan Digital Sky Survey V has been provided by the Alfred P. Sloan Foundation, the Heising-Simons Foundation, the National Science Foundation, and the Participating Institutions. SDSS acknowledges support and resources from the Center for High-Performance Computing at the University of Utah. The SDSS web site is \url{www.sdss.org}.

SDSS is managed by the Astrophysical Research Consortium for the Participating Institutions of the SDSS Collaboration, including the Carnegie Institution for Science, Chilean National Time Allocation Committee (CNTAC) ratified researchers, the Gotham Participation Group, Harvard University, Heidelberg University, The Johns Hopkins University, L’Ecole polytechnique f{\'e}d{\'e}rale de Lausanne (EPFL), Leibniz-Institut f{\"u}r Astrophysik Potsdam (AIP), Max-Planck-Institut f{\"u}r Astronomie (MPIA Heidelberg), Max-Planck-Institut f{\"u}r Extraterrestrische Physik (MPE), Nanjing University, National Astronomical Observatories of China (NAOC), New Mexico State University, The Ohio State University, Pennsylvania State University, Smithsonian Astrophysical Observatory, Space Telescope Science Institute (STScI), the Stellar Astrophysics Participation Group, Universidad Nacional Aut{\'o}noma de M{\'e}xico, University of Arizona, University of Colorado Boulder, University of Illinois at Urbana-Champaign, University of Toronto, University of Utah, University of Virginia, Yale University, and Yunnan University.

The Pan-STARRS1 Surveys (PS1) and the PS1 public science archive have been made possible through contributions by the Institute for Astronomy, the University of Hawaii, the Pan-STARRS Project Office, the Max-Planck Society and its participating institutes, the Max Planck Institute for Astronomy, Heidelberg and the Max Planck Institute for Extraterrestrial Physics, Garching, The Johns Hopkins University, Durham University, the University of Edinburgh, the Queen's University Belfast, the Harvard-Smithsonian Center for Astrophysics, the Las Cumbres Observatory Global Telescope Network Incorporated, the National Central University of Taiwan, the Space Telescope Science Institute, the National Aeronautics and Space Administration under Grant No. NNX08AR22G issued through the Planetary Science Division of the NASA Science Mission Directorate, the National Science Foundation Grant No. AST-1238877, the University of Maryland, Eotvos Lorand University (ELTE), the Los Alamos National Laboratory, and the Gordon and Betty Moore Foundation.
\section*{Data Availability}
The LAMOST DR9 spectra are available on LAMOST data archive \href{http://www.lamost.org/lmusers/}{http://www.lamost.org/lmusers/}. The SDSS data are available at \href{https://www.sdss4.org/dr17/data_access/}{https://www.sdss4.org/dr17/data\_access/}. The Pan-STARR data are available at \href{https://outerspace.stsci.edu/display/PANSTARRS}{https://outerspace.stsci.edu/display/PANSTARRS}. The WISE data can be available at \href{http://wise.ssl.berkeley.edu}{http://wise.ssl.berkeley.edu}. The VLA-FIRST data are available at \href{http://first.astro.columbia.edu}{http://first.astro.columbia.edu}. 
The GALEX data access via \href{https://archive.stsci.edu/missions-and-data/galex}{https://archive.stsci.edu/missions-and-data/galex}.
We use python package {\it astroquery} \citep{Ginsburg2019} to query data in each database and use python package {\it Astropy} \footnote{https://www.astropy.org} to merge tables. We use python package {\it pyspeckit}\footnote{https://pyspeckit.readthedocs.io} for spectral analyses.



\bibliographystyle{mnras}
\bibliography{example} 

\bsp	
\label{lastpage}
\end{document}